\documentclass[prd, aps,
superscriptaddress,showpacs,preprintnumbers,floatfix,
nofootinbib,twocolumn]{revtex4-1} 
\usepackage[utf8]{inputenc}
\usepackage{epsfig}
\usepackage{graphicx}
\usepackage{graphics}
\usepackage{pstricks,pstricks-add}
\usepackage{pst-plot}
\usepackage{epstopdf} 
\usepackage{amsthm} 
\usepackage{amsfonts}
\usepackage{amsmath}
\usepackage{amssymb}
\usepackage{bm}
\usepackage{dcolumn}
\usepackage{latexsym}
\usepackage{rotating}
\usepackage{hyperref}
\usepackage{color}

\definecolor{linkcolor}{rgb}{0.0,0.3,0.5}
\definecolor{venetianred}{rgb}{0.78, 0.03, 0.08}
\hypersetup{colorlinks=true,linkcolor=linkcolor,citecolor=linkcolor,filecolor=linkcolor,urlcolor=linkcolor}

\newcommand{\dd}{\mathrm d}
\newcommand{\ex}{\mathrm e}

\newcommand{\om}{\ensuremath{\omega}}

\newcommand{\pd}{\ensuremath{\partial}}
\newcommand{\la}{\ensuremath{\lambda}}
\newcommand{\abs}[1]{\ensuremath{\left \lvert #1 \right\rvert}} 
\newcommand{\ii}{\ensuremath{\mathrm{i}}}
\newcommand{\e}{\ensuremath{\mathrm{e}}}
\newcommand{\s}{\nobreak\hspace{.08em plus .04em}}
\newcommand{\lp}{\ensuremath{\left(}}
\newcommand{\rp}{\ensuremath{\right)}}
\newcommand{\lc}{\ensuremath{\left[}}
\newcommand{\rc}{\ensuremath{\right]}}
\newcommand{\lb}{\ensuremath{\left\lbrace}}

\newcommand{\fz}{0.9\linewidth} 
\newcommand{\fzz}{0.495\linewidth} 

\newcommand{\A}{B}
\newcommand{\B}{A}
\newcommand{\F}{G}
\newcommand{\G}{F}

\begin{document}

\title{Excited cosmic strings with superconducting currents}

\author{Betti Hartmann}
\affiliation{Instituto de F\'isica de S\~ao Carlos (IFSC), Universidade
de S\~ao Paulo (USP), CP 369,  13560-970 , S\~ao Carlos, SP, Brasil}

\author{Florent Michel}
\affiliation{Centre for Particle Theory, Durham University, South Road,
Durham, DH1 3LE, UK}

\author{Patrick Peter}
\affiliation{Institut d'Astrophysique de Paris ($\mathcal{G} \mathbb{R}
\varepsilon \mathbb{C} \mathcal{O}$), UMR 7095 CNRS, Sorbonne
Universités, UPMC Univ. Paris 06, Institut Lagrange de Paris, 98 bis
boulevard Arago, 75014 Paris, France.}

\date{\today}

\begin{abstract}
We present a detailed analysis of {\it excited} cosmic string solutions
which possess superconducting currents. These currents can be excited
inside the string core, and -- if the condensate is large enough -- can
lead to the excitations of the Higgs field. Next to the case with global
unbroken symmetry, we discuss also the effects of the gauging of this
symmetry and show that excited condensates persist when coupled to an
electromagnetic field. The space-time of such strings is also
constructed by solving the Einstein equations numerically and we show
how the local scalar curvature is modified by the excitation. We
consider the relevance of our results on the cosmic string network
evolution as well as observations of primordial gravitational waves and
cosmic rays.
\end{abstract}

\maketitle

\section{Introduction}

Although cosmic strings
\cite{Nielsen:1973cs,Kibble:1976sj,Vilenkin:1984ib,Hindmarsh:1994re,Vilenkin:2000jqa,Vachaspati:2015cma}, 
i.e. linear topological defects expected to have formed at phase
transitions during the early stages of the Universe, are no longer
accepted as candidates for cosmic microwave background (CMB) primordial
fluctuations \cite{Ade:2015xua} (See Ref.~\cite{Ade:2013xla} for an
update on the cosmic string search in the CMB and the more recent work
\cite{Ciuca:2017gca} in which new methods are being developed), they are
still expected to be produced in the grand unified theory (GUT)
framework (see, e.g., Ref.~\cite{Allys:2015kge} and references therein),
in which case they are very likely to have bosonic condensates
\cite{Allys:2015yda} or be current-carrying \cite{Davis:1995kk}. The
structure of such objects has been studied in detail for many models,
from the original Witten \cite{Witten:1984eb} fermionic
\cite{Ringeval:2000kz,Ringeval:2001xd} or bosonic kind
\cite{Babul:1987me,Peter:1992dw,Peter:1992ta,Peter:1993mv}, leading to
effective equations of state \cite{Carter:1994hn,Hartmann:2008yr}
potentially useful for large scale network simulations
\cite{Ringeval:2012tk,Rybak:2017yfu}. Until the reason why strings have
yet not been observed in the CMB is clarified, it is therefore of utmost
importance to understand in as many details as possible their internal
structure and the associated plausible cosmological consequences.

In a previous work \cite{Hartmann:2016axn}, by investigating the neutral
current-carrying Witten model \cite{Peter:1992dw}, we identified a new
set of excited solutions in which the condensate oscillates and thus
yields a many-valued equation of state, i.e. we found several (possibly
many, depending on the parameters) different branches in the energy per
unit length and tension as functions of the state parameter. We also
argued that those new modes should be unstable and deduced some
plausible cosmological consequences. The purpose of this work is to
deepen our understanding of these modes and to make the argument for
their instability more rigorous. We also discuss inclusion of
electromagnetic-like effects \cite{Peter:1992ta,Peter:1993mv} if the
current is coupled to a massless gauge field. Finally, we couple our
model to gravity in order to derive the local
\cite{Babul:1988qt,Amsterdamski:1988zp,Peter:1993ww} and asymptotic
\cite{Moss:1987ka,Linet:1989xj,Peter:1993wx,Christensen:1999wb,Brihaye:2000qr,Hartmann:2012mh} geometrical structure.

An interesting new outcome of this detailed investigation is that the
string-forming Higgs field itself may oscillate in a restricted regime
of parameter space, which leads to oscillations in the gravitational
field around the vortex, thus potentially enhancing the gravitational
waves produced by a network of such strings and leading to the emission
of high energy particles.

Besides their possible relevance for cosmology, these solutions may have
close analogues in atomic Bose-Einstein condensates. Indeed, it is now
well known (see for instance~\cite{2001JPCM...13R.135F} and references
therein) that one-dimensional vortex lines can arise in rotating
condensates. Considering a dilute gas of two types of atoms with
different transition frequencies, it should be possible to tune the
potential to mimic the Higgs field-condensate interactions in
superconducting strings. One would then expect solutions with a similar
structure and basic properties, although the stability analysis would be
somewhat different since non-relativistic condensates obey a first-order
equation in time, so that, in particular, the analogues of the unstable 
modes with imaginary frequencies found in Section~\ref{sub:stability}
would be negative-energy modes in the non-relativistic case. Such
analogies between cosmological phenomena and condensed-matter systems
have been fruitful in the context of black-hole
physics~\cite{Unruh:1980cg, Barcelo:2005fc, Weinfurtner:2010nu,
Euve:2015vml, Steinhauer:2015saa}, in particular clarifying the effects
of Lorentz violations on Hawking radiation~\cite{Jacobson:1991gr} and
leading to the discovery of new phenomena in condensed-matter systems.
It is conceivable that a detailed study of such excited vortex lines in
condensed matter would also reveal new interesting physics.

The purpose of this paper is to  detail and complement the results of
the analysis of Ref.~\cite{Hartmann:2016axn}, in which the
electromagnetic-like U(1) symmetry of the model  was in fact not gauged,
thus corresponding to neutral currents flowing along the string
\cite{Peter:1992dw}. This is done in Sec. \ref{sec:neutral}. In
particular, we present new results related to the back-reaction of the
excited condensate on the Higgs field.

The effects due to a nonvanishing value of the
electromagnetic-like\footnote{According to the standard model of
particle physics however, such a massless U(1) gauge boson corresponds
unambiguously to the photon and the relevant symmetry to that of actual
electromagnetism. We keep referring to an electromagnetic-like coupling
because the structure we are investigating here might be only temporary,
with the symmetry being only unbroken as an intermediate step in a full
GUT symmetry-breaking scheme leading to the standard model.} coupling
are discussed briefly in Sec.~\ref{sec:gauged} and the gravitational
effects are presented in Sec.~\ref{sec:GE}. In Sec.~\ref{Conclusions} we
discuss our results and conclude.

\section{The model}

The underlying toy model describing a current-carrying vortex
(superconducting cosmic string) has been proposed by Witten in 1985
\cite{Witten:1984eb}. It consists in two complex scalar fields $\phi$
and $\sigma$, each subject to independent phase shift invariance, both
of which being possibly gauged. The general situation is therefore the
so-called U(1)$\times$U(1) scalar Witten model, which reads
\begin{align}
\mathcal{L}=&\frac{1}{2} (D_{\mu} \phi)(D^{\mu} \phi)^{*} +\frac{1}{2}
(D_{\mu} \sigma)(D^{\mu} \sigma)^{*} -V(\phi,\sigma) \nonumber \\ &
-\frac{1}{4} \F_{\mu\nu}\F^{\mu\nu}-\frac{1}{4} \G_{\mu\nu}\G^{\mu\nu}.
\label{lag}
\end{align}
Here $\F_{\mu\nu}$ and $\G_{\mu\nu}$ denote the field strength tensors
of the two $\mathrm{U}(1)$ gauge fields $\A_{\mu}$ and $\B_{\mu}$
respectively, namely
\begin{equation}
\F_{\mu\nu}=\partial_{\mu} \A_{\nu} - \partial_{\nu} \A_{\mu} \ \
\hbox{and} \ \ \G_{\mu\nu}=\partial_{\mu} \B_{\nu} - \partial_{\nu}
\B_{\mu},
\end{equation}
and the covariant derivatives read
\begin{equation}
D_{\mu}\phi =\partial_{\mu} \phi - \ii \s e_1 \phi \A_{\mu} \ \
\hbox{and}  \ \ D_{\mu}\sigma =\partial_{\mu} \sigma - \ii \s e_2 \sigma
\B_{\mu},
\end{equation}
where $e_1$ and $e_2$ are the coupling constants of the respective
scalar fields $\phi$ and $\sigma$ to the corresponding gauge fields.
Finally, we set the potential to
\begin{equation}
V= \frac{\lambda_1}{4}
\left(\vert\phi\vert^2-\eta_1^2\right)^2
+\frac{\lambda_2}{4}\vert\sigma\vert^2
\left(\vert\sigma\vert^2-2\eta_2^2\right)
+\frac{\lambda_3}{2}\vert\phi\vert^2 \vert\sigma\vert^2, \label{pot}
\end{equation}
which is the most general renormalizable one given the field content.

In what follows, we choose the parameters of the potential \eqref{pot}
above in such a way that the $\mathrm{U}(1)$ symmetry associated to the
fields $\phi$ and $\A_{\mu}$ gets spontaneously broken, thereby forming
an Abelian-Higgs string, while the $\mathrm{U}(1)$ symmetry associated
to the fields $\sigma$ and $\B_{\mu}$ remains unbroken. Associated to
this unbroken symmetry the cosmic string will carry a locally conserved
Noether current and a globally conserved Noether charge, which in the
gauged case can be interpreted as electromagnetic current and charge,
respectively. 

\subsection{Field equations}

The ansatz for the vector fields in cylindrical coordinates
$(r,\theta,z)$ reads
\begin{align} 
\A_{\mu} \dd x^{\mu} = &\frac{1}{e_1} \left[n - P(r)\right] \dd \theta, \nonumber \\
\B_{\mu} \dd x^{\mu} = &\frac{1-b(r)}{e_2} \left(\omega \dd t - k \dd z\right), 
\label{eq:anz1}
\end{align}
while the scalar fields take the form
\begin{equation} \label{eq:anz2}
\phi(r,\theta,z)=\eta_1 h(r)\ex^{\ii n\theta} \ \ , \ \
\sigma(r,\theta,z)=\eta_1 f(r)\ex^{\ii (\omega t-kz)}
\end{equation}

We introduce the following dimensionless coordinate and energy ratio
\begin{equation}
x \equiv  \sqrt{\lambda_1} \eta_1 r, \ \ \ \ \ q=\frac{\eta_2}{\eta_1},
\end{equation}
and the rescaled coupling constants
\begin{equation}
\alpha_i^2=\frac{e_i^2}{\lambda_1}, \ \ \ \ \hbox{and} \ \ \ \ \ 
\gamma_i = \frac{\lambda_i}{\lambda_1} \ \ \ \  (i=2,3).
\end{equation}
We also rescale the Lagrangian into the dimensionless quantity
$\mathcal{L} \rightarrow \tilde{\mathcal{L}} := \mathcal{L}/(\lambda_1
\eta_1^2)$.

With these notations, the equations of motion read
\begin{align}
\left(\frac{P'}{x}\right)' & = \alpha_1^2 \frac{P h^2}{x},
\label{gauge} \\
\frac{1}{x} (xb')' & =\alpha_2^2 b f^2,
\label{gauge2} \\
\frac{1}{x}\left( x h'\right)' & = \frac{P^2 h }{x^2}  + h (h^2 - 1) + \gamma_3 f^2  h ,
\label{higgs} \\
\frac{1}{x}\left( x f'\right)' & = \tilde w f b^2 + \gamma_2 f (f^2 - q^2) +
\gamma_3 f  h^2,
\label{condensate}
\end{align}
where a prime denotes a derivative with respect to $x$ and we have
defined the state parameter $w$ as $w:=k^2-\omega^2 = \lambda_1 \eta_1^2
\tilde w$, thereby defining its rescaled counterpart $\tilde w$. The
sign of the state parameter $w$ is defined positive for a spacelike
current ($w>0$) and negative for a timelike current ($w<0$), while $w=0$
corresponds to a chiral (lightlike) current.

The necessary boundary conditions corresponding to a current-carrying vortex
then read
\begin{equation}
P(0)=n, \ b(0)=1, \ h(0)=f'(0)=b'(0)=0,\label{Bound1}
\end{equation}
at the origin and
\begin{equation}
\lim_{x\to\infty} P(x)=\lim_{x\to\infty} \sqrt{x} f(x)=0, \ \ \hbox{and} \ \
\lim_{x\to\infty}h(x)=1, \label{Bound2}
\end{equation}
at infinity. Although we have produced solutions with $n > 1$ which we
briefly comment upon in Sec.~\ref{sub:largern}, for the most part of the
following, we work with $n=1$ for definiteness.

\subsection{Integrated quantities}

Cosmological consequences of the existence of topological defects can be
studied under the approximation that they are infinitely thin in their
transverse dimension compared with their longitudinal extension. This
amounts to integrating over the transverse dimensions. In our case, the
relevant quantities are the energy per unit length $U$ and tension
$T$. Those are calculated as the eigenvalues of the integrated 
stress-energy tensor\footnote{Note that there is a degeneracy in the
structureless (currentless) case leading to the usual Nambu-Goto action
for which $U=T$.}
\begin{equation}
\bar T_{\mu\nu} := \int \lp -2 \frac{\delta\mathcal{L}}{\delta
g^{\mu\nu}} +g_{\mu\nu} \mathcal{L}\rp \dd^2 x^\perp ,
\label{TmunuDef}
\end{equation}
where in the present symmetric situation the relevant integration
measure element across the string is given by $\int_{\theta = 0}^{2 \pi}
\dd^2 x^\perp = \int_{\theta = 0}^{2 \pi} r\dd r \dd\theta = 2\pi \s r
\s \dd r$. To figure them, we restrict attention to the worldsheet space
coordinates $\xi^a \in \{ t, z\}$, i.e., we explicit the matrix $\bar
T^{ab}$ and find the eigenvalues by solving the characteristic equation
$\det\lp \bar T^{ab} - \lambda \eta^{ab}\rp =0$, with the 2-dimensional
Minkowski metric $\eta^{ab} := \mathrm{diag}\, \{1,-1\}$. This leads to
\begin{equation}
\lp \begin{matrix} U\cr T \end{matrix} \rp = \eta_1^2
\lp \begin{matrix} \tilde U
\cr \tilde T \end{matrix}
\rp = \pi\eta_1^2 \int \lp \sum_{i=1}^3 \varepsilon_i \pm c +u \rp\,
x\,\dd x, \label{UT}
\end{equation}
where
\begin{eqnarray}
\varepsilon_1 &:=& h'^2 + f'^2, \\
\varepsilon_2 &:=& \frac{h^2 P^2}{x^2}, \\
\varepsilon_3 &:=& \frac{P'^2}{\alpha_1^2 x^2},\\
c &:=& |\tilde w| \lp \frac{b'^2}{\alpha_2^2} + f^2 b^2 \rp , \\
u &:= & \frac{1}{2} (h^2-1)^2 + \frac{\gamma_2}{2} f^2 (f^2 - 2 q^2) +
\gamma_3 h^2 f^2.
\end{eqnarray}
This form clearly makes all the relevant quantities Lorentz invariant;
in Eq.~\eqref{UT}, the meaning of the column vector is that $U$
corresponds to the $+$ sign in front of the quantity $c$, while $T$ is
calculated with the $-$ sign (this ensures that $U\geq T$). These
definitions of $U$ and $T$ are valid even in the electromagnetically
coupled case $e_2\not=0$, even though we mostly concentrate in what
follows on the neutral case $e_2=0$.

The velocities of longitudinal and transversal perturbations which are
given by $c_\mathrm{_L}=\sqrt{-\dd T/\dd U}$ and
$c_\mathrm{_T}=\sqrt{T/U} $, respectively, should both be real in order
for the string to be stable \cite{Carter:1989xk}. This requires $T/U >
0$ and $\dd T/\dd U  < 0$, conditions which we refer to below as
Carter stability conditions.

Another quantity of interest is the current flowing along the
worldsheet. Starting from the U(1) invariance of $\sigma$, one forms the
microscopic current
\begin{equation}
J^\mu := \frac{1}{e_2}\frac{\delta \mathcal{L}}{\delta \B_\mu} = - \eta_1^2 f^2 
\lc \partial^\mu \lp \omega t -kz\rp - e_2 \B^\mu\rc,
\label{Jmic}
\end{equation}
where the normalizing factor $1/e_2$ ensures it remains finite in the
neutral limit $e_2\to 0$. Integrating radially again yields the current
$C$
\begin{equation}
C:= \int\dd^2 x^\perp \sqrt{\left| \eta_{ab} J^a  J^b \right|}. 
\label{Jtilde}
\end{equation}
This gives explicitly, in terms of the field functions 
\begin{equation}
C = 2\pi |v| \eta_1^2 \int f^2 b\, r \,\dd r = 2\pi 
\frac{\eta_1}{\sqrt{\lambda_1}} \tilde{C}, \label{C}
\end{equation}
where the reduced state parameter is $v =
\mathrm{sign}(w)\sqrt{|w|}=\sqrt{\lambda_1} \eta_1 \tilde v$; the
meaning of this parameter is clear: for a spacelike current, there
exists a frame in which $\omega\to 0$ and $w\to k^2$, in which case
$v \to k$, while for a timelike current, there exists a frame where
$k\to 0$, so that $v \to-\omega$ (the sign is included 
in order to clearly distinguish between spacelike and timelike
configurations and for convenience when it comes to plotting).

\section{Solutions in the neutral model}
\label{sec:neutral}

In the following, we will concentrate on the case $\alpha_2=0$, i.e. the
case in which the current along the string is ungauged, which implies
$b(x)\equiv 1$.

\subsection{Linear condensate} 

To motivate the existence of excited solutions, we work in a regime
where the condensate is sufficiently small to neglect its backreaction
on the string-forming Higgs scalar $h$. To reduce the number of
parameters, we define the shifted squared frequency $\Omega \equiv
\tilde w - \gamma_2 q^2$. Then, Eq.~\eqref{condensate} becomes
\begin{equation}
f'' + \frac{1}{x} f'  = \lp \Omega + \gamma_3 h^2 \rp f + \gamma_2 f^3.
\label{eq:f2}
\end{equation}
We look for ``bound state'' solutions which are regular at $x = 0$, not
equal to zero everywhere (i.e., we discard the trivial solution $f =
0$), and decay strictly faster than $x^{-1/2}$ at
infinity.\footnote{This condition ensures that there is no quadratic
conserved flux at infinity, in accordance with the usual definition of a
bound state.} One can obtain two bounds on $\Omega$, namely
\begin{equation}
-\gamma_3 \leq \Omega < 0 \ \ \ \Longrightarrow - m_\sigma^2
\leq w < M_\sigma^2,
\label{OmegaBound}
\end{equation}
where $m_\sigma^2 := \lp \lambda_3 \eta_1^2 - \lambda_2 \eta_2^2\rp$ is
the rest mass of the current carrier $\sigma$ field outside the string
where $|\phi|\to \eta_1$, and $M_\sigma^2:= \lambda_2 \eta_2^2$ its mass
inside the string where $\phi\to 0$. The first bound, first obtained in
Ref.~\cite{Peter:1992dw}, shows there exists a phase frequency
threshold; it merely reflects the fact that it is energetically favored
for a trapped particle with energy larger than its asymptotic mass to
flow away from the string core. They are obtained through the following
arguments:
\begin{itemize}
\item If $\Omega \geq 0$, since $\gamma_3>0$, $f'' + \frac{1}{x} f'$ has
everywhere the same sign as $f$. Assume first $f(0)>0$. Since $\lc x
f'(x)\rc ' > 0$ for sufficiently small $x > 0$ and (obviously) $x f'(x)
= 0$ at $x = 0$, this implies that $\lc x f'(x)\rc  > 0$, and therefore
that $f'(x) > 0$, for sufficiently small positive values of $x$: the
function $f$ thus grows. Therefore, in order for $f$ to vanish
asymptotically, it must stop growing at some stage, and hence it must go
though a maximum: $\exists x_\mathrm{max}; f'(x_\mathrm{max}) = 0 \ \& \
f''(x_\mathrm{max}) \leq0$. But we also have, by construction, that
$f(x_\mathrm{max})>0$, implying $f''(x_\mathrm{max})+\frac{1}{x}
f'(x_\mathrm{max}) =f''(x_\mathrm{max})>0$, in contradiction with the
hypothesis. The function $f(x)$ thus grows indefinitely. For $f(0)<0$,
the same argument applies in the opposite direction, showing that $f(x)$
decreases for all values of $x$, while the case $f(0) = 0$ would lead to
the trivial solution $f(x) = 0$ for all $x$. As a result, we deduce that
$\forall x > 0, \left\lvert f(x) \right\rvert \geq \left\lvert f(0)
\right\rvert$. This is clearly in contradiction with the assumption that
$f$ goes to zero at infinity, so we must set $\Omega <0$.

\item Let us now show that $\Omega \geq - \gamma_3$. To this end, it is
convenient to define the function $s(x) := \sqrt{x} f(x)$.
Eq.~\eqref{eq:f2} may be rewritten as
\begin{equation} \label{eq:ints} s'' +
\frac{s}{4 x^2} = \lp \Omega + \gamma_3 h^2 \rp s + \frac{\gamma_2}{x}
s^3.
\end{equation}
To simplify the notations, let us also define the two quantities $K := -
\lp \Omega + \gamma_3 \rp$ and
\begin{equation}
\Theta(x) := -\frac{1}{4 x^2} + \gamma_3 \left[ h(x)^2 - 1 \right] +
\frac{\gamma_2}{x} s(x)^2,
\end{equation}
in terms of which Eq.~\eqref{eq:ints} becomes
\begin{align} 
s''(x) = - K \s s(x) + \Theta(x) \s s(x),
\end{align}
which gives, upon multiplication by $2 \s s'(x)$ on both sides,
\begin{align}
\label{eq:enes}
\frac{\dd}{\dd x} \lp s'^2 + K \s s^2 \rp = 2 \s \Theta \s s' \s s.
\end{align}
Eq.~\eqref{eq:enes} is our main tool to prove the desired result.
Indeed, as we now show, if $K > 0$, i.e., $\Omega < - \gamma_3$, then
the ``energy'' $s'^2 + K \s s^2$ does not go to zero at infinity, in
contradiction with the definition of a localized state.

For clarity, let us list explicitly the properties of the functions $s$
and $h$ we will use. First, we assume that $s$ is not identically zero,
i.e., that a condensate is present inside the string. Second, we use
that $h$ and $f$, and thus $s$, converge to zero exponentially at
infinity, as shown in~\cite{Peter:1992dw}. This implies that
\begin{enumerate}
\item $h^2 - 1$ is integrable on the interval $x \in [0 , + \infty[$;
\item $s^2 / x$ is integrable on the interval $x \in [1 , + \infty[$;
\item $s'(x)$ goes to zero as $x \to \infty$.
\end{enumerate}
The function $\Theta$ is thus absolutely integrable at infinity. If $K
\neq 0$, there thus exists $x_1 > 0$ such that\footnote{There is, of
course, an infinite number of possible choices: any sufficiently large
value of $x_1$ will satisfy this property.}
\begin{align}\label{eq:intTheta}
\int_{x_1}^\infty \abs{\Theta(x)} \dd x < \frac{\sqrt{\abs{K}}}{2}. 
\end{align}
This is the crucial point, which allows us to bound the variation of the
``energy'' $s'^2 + K \s s^2$.

We now have all the elements to prove the desired result. As in the
first point, we proceed by contradiction. Let us assume that $K > 0$ and
define $M_\mathrm{s} \equiv {\rm sup}_{x > x_1} \abs{s' \, s}$. Since
$s$ is not a constant function, $s \s s'$ takes nonvanishing values, so
$M_\mathrm{s} > 0$. Moreover, since we demand that $s(x)$ and $s'(x)$
must vanish asymptotically, $\abs{s(x) \s s'(x)}$ goes to zero in this
limit, so $M_\mathrm{s}$ is reached at some point $x_2 \geq x_1$. Using
that $\lc s'(x_2) \pm \sqrt{K} \s s(x_2)\rc^2 \geq 0$, one obtains
\begin{align}\label{eq:int1}
s'(x_2)^2 + K \s s(x_2)^2 & \geq 2 \sqrt{K} \s \abs{s'(x_2) \s s(x_2)}
\nonumber \\ & \geq 2 \sqrt{K} \s M_\mathrm{s}.
\end{align}
On the other hand, from Eq.~\eqref{eq:enes},
\begin{align}
\hskip1cm \abs{\int_{x_2}^\infty \frac{\dd}{\dd x} \lp s'^2 + K s^2 \rp \dd x} & = 
\abs{\int_{x_2}^\infty 2 \Theta s' s\, \dd x}\nonumber\\
& \hskip-1cm \leq
2 M_\mathrm{s} \int_{x_2}^\infty \abs{\Theta} \dd x <
\sqrt{K} \s M_\mathrm{s},
\end{align}
where Eq.~\eqref{eq:intTheta} was used in the last step. 
We thus have:
\begin{align}\label{eq:int2}
\left[ (s')^2 + K \s s^2 \right]_{x_2}^\infty < \sqrt{K} \s M_\mathrm{s}. 
\end{align}
Combining Eqs.~\eqref{eq:int1} and~\eqref{eq:int2}, we deduce that
$$\lim_{x\to\infty} \left[ s'(x)^2 + K \s s^2(x) \right] > \sqrt{K} M_\mathrm{s},$$ 
in contradiction with the assumption that $s$ and $s'$ both go to zero in this limit. 
We conclude that solutions can exist only if $K \leq 0$, i.e., if $\Omega \geq - \gamma_3$.
\end{itemize}

In order to motivate the existence of our excited modes, we further
assume that the nonlinear term in \eqref{condensate} is negligible, and
we work with the following simple continuous but non differentiable
ansatz for the function $h$:
\begin{align}
h(x) = \left\lbrace
\begin{array}{ll}
\kappa \, x & \ \ \hbox{for} \ \ 0 \leq x < 1 / \kappa, \\ 
1 & \ \ \hbox{for} \ \  x> 1 / \kappa.
\end{array} 
\right.
\end{align}
This simple form provides a strong motivation for the existence of
excited solutions and allows to determine some of their expected
properties.

For $x \geq 1/\kappa$, $f$ satisfies a modified Bessel
equation~\cite{Olver:2010:NHM:1830479}. The only solutions going to zero
sufficiently fast at infinity are
\begin{align}
f(x) = C_1 K_0 \left( \sqrt{\Omega + \gamma_3} x \right), \, C_1 \in \mathbb{R}.
\end{align}
To solve the equation in the interior region $x < 1/\kappa$, it is
useful to define the variable $Y \equiv \sqrt{\gamma_3} \kappa x^2$
and the function $F$ by $f(x) = \exp[-Y(x)/2] F[Y(x)]$. Doing this, we
obtain
\begin{align}
Y F'' + (1-Y) F' + \mathcal{A} \, F = 0, \; \mathcal{A} \equiv -
\left( \frac{\Omega}{4 \sqrt{\gamma_3} \kappa} + \frac{1}{2} \right).
\end{align}
This is the confluent hypergeometric
equation~\cite{Olver:2010:NHM:1830479}. The only regular solutions are
$F(Y) \propto L_\mathcal{A}(Y)$, where $L_\mathcal{A}$ denotes the
Laguerre function with parameter $\mathcal{A}$. So, for $x > 1/\kappa$,
\begin{align}
f(x) = C_2 \e^{- \sqrt{\gamma_3} \kappa x^2/2}
L_\mathcal{A}(\sqrt{\gamma_3} \kappa x^2), \, C_2 \in \mathbb{R}.
\end{align}
Since \eqref{eq:f2} has no singularity at $x = 1/\kappa$, $f$ and $f'$
must be continuous at that point, and this provides two matching
conditions. A straightforward calculation shows they can be
simultaneously satisfied if and only if

\begin{equation}
2 \frac{L_{\mathcal{A}}'}{L_{\mathcal{A}}} = 1+
\frac{K'_0 \lp \sqrt{\mu+1} \xi \rp \sqrt{\mu+1}}{K_0\lp \sqrt{\mu+1} \xi \rp},
\label{eq:matching}
\end{equation}
where $\mu \equiv \Omega / \gamma_3$ and $\xi \equiv \sqrt{\gamma_3} /
\kappa$. Then, $C_1$ and $C_2$ are related through
\begin{align}
\frac{C_1}{C_2} = \frac{L_{\mathcal{A}}\left( \sqrt{\gamma_3} / \kappa
\right)}{K_0 \left( \sqrt{\Omega + \gamma_3} / \kappa \right)}
\e^{-\sqrt{\gamma_3} \kappa/2}.
\end{align}

\begin{figure}
\begin{center}
\includegraphics[width=\fz]{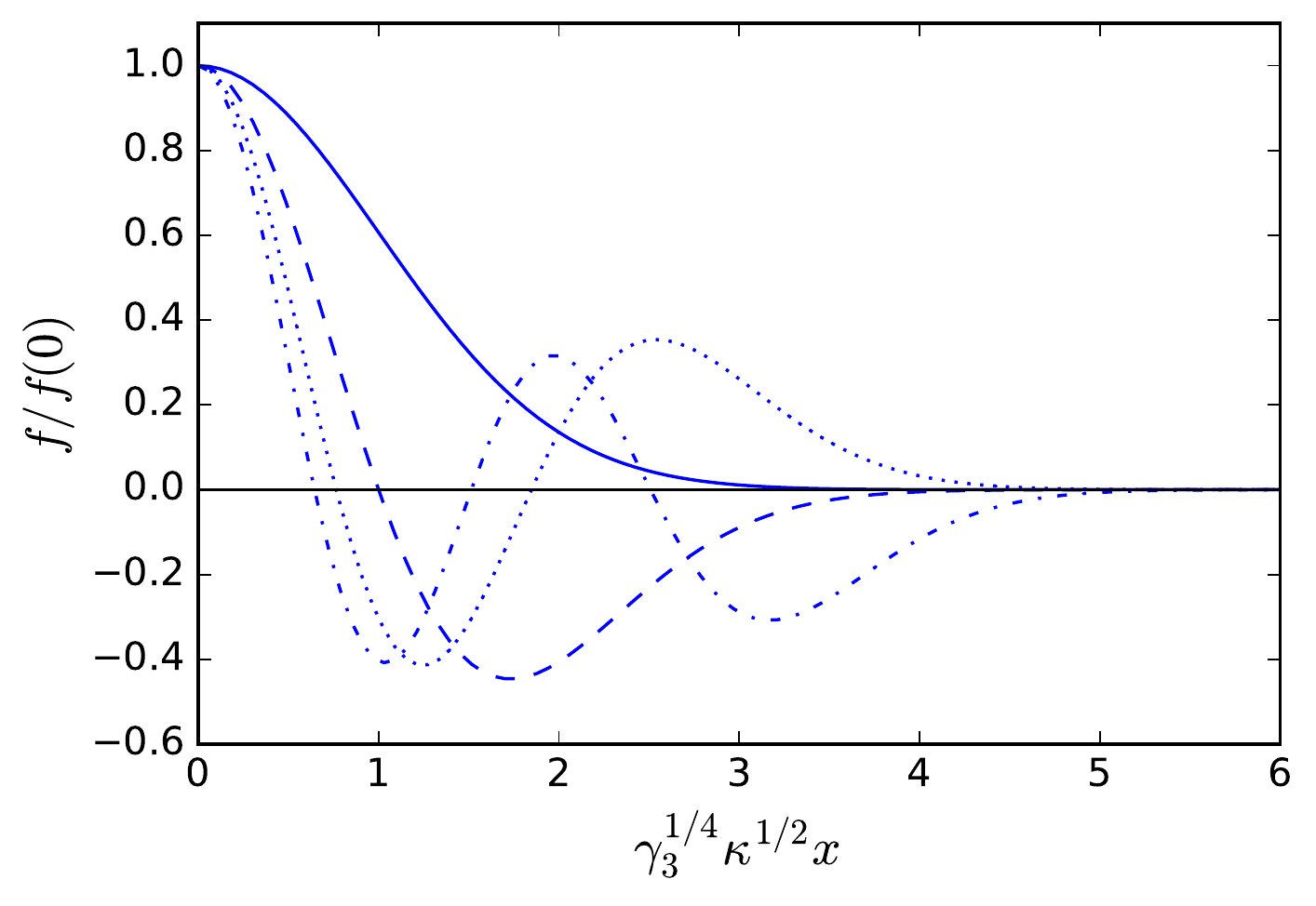}
\end{center}
\caption{Fundamental (continuous) and first three excited (dashed for
$m = 1$, dotted for $m = 2$, and dot-dashed for $m = 3$) solutions in
the limit $\xi \gg 1$.}
\label{fig:Laguerre1}
\end{figure}

To our knowledge, \eqref{eq:matching} can not be solved analytically in
general. However, it greatly simplifies in the limit $\xi \gg 1$, i.e.,
for very small $\kappa$. Then $f$ is negligible for $x > 1/\kappa$ and
is approximately given by a globally regular solution of the Laguerre
equation going to zero at infinity. The latter are the Laguerre
polynomials~\cite{Olver:2010:NHM:1830479}, which exist if and only if
$\mathcal{A} \in \mathbb{N}$. Moreover, the $n^{\rm th}$ Laguerre
polynomial has $n-1$ strictly positive roots. The corresponding solution
in $f$ thus has $m = n-1$ nodes. The four first ones are shown in
Fig.~\ref{fig:Laguerre1}.
This approximation is valid provided all nodes are well inside the
interior region, i.e., $m \ll \xi / 4$. We thus expect that solutions
with $m$ nodes exist up to a maximum value $m_\mathrm{max}$ close to
$\xi / 4$. One can also estimate the positions of the roots using the
explicit form of the Laguerre polynomials. For instance, for the first
excited solution, we find that the unique root is at
\begin{equation}
x_0 \approx \left( \gamma_3 \kappa^2 \right)^{-1/4},
\label{zero}
\end{equation}
while for the second excited solution, the two roots are at $x_0 \approx
\left[ (2 \pm \sqrt{2}) / \sqrt{\gamma_3 \kappa^2} \right]^{-1/2}$.

\begin{figure}
\includegraphics[width=\fz]{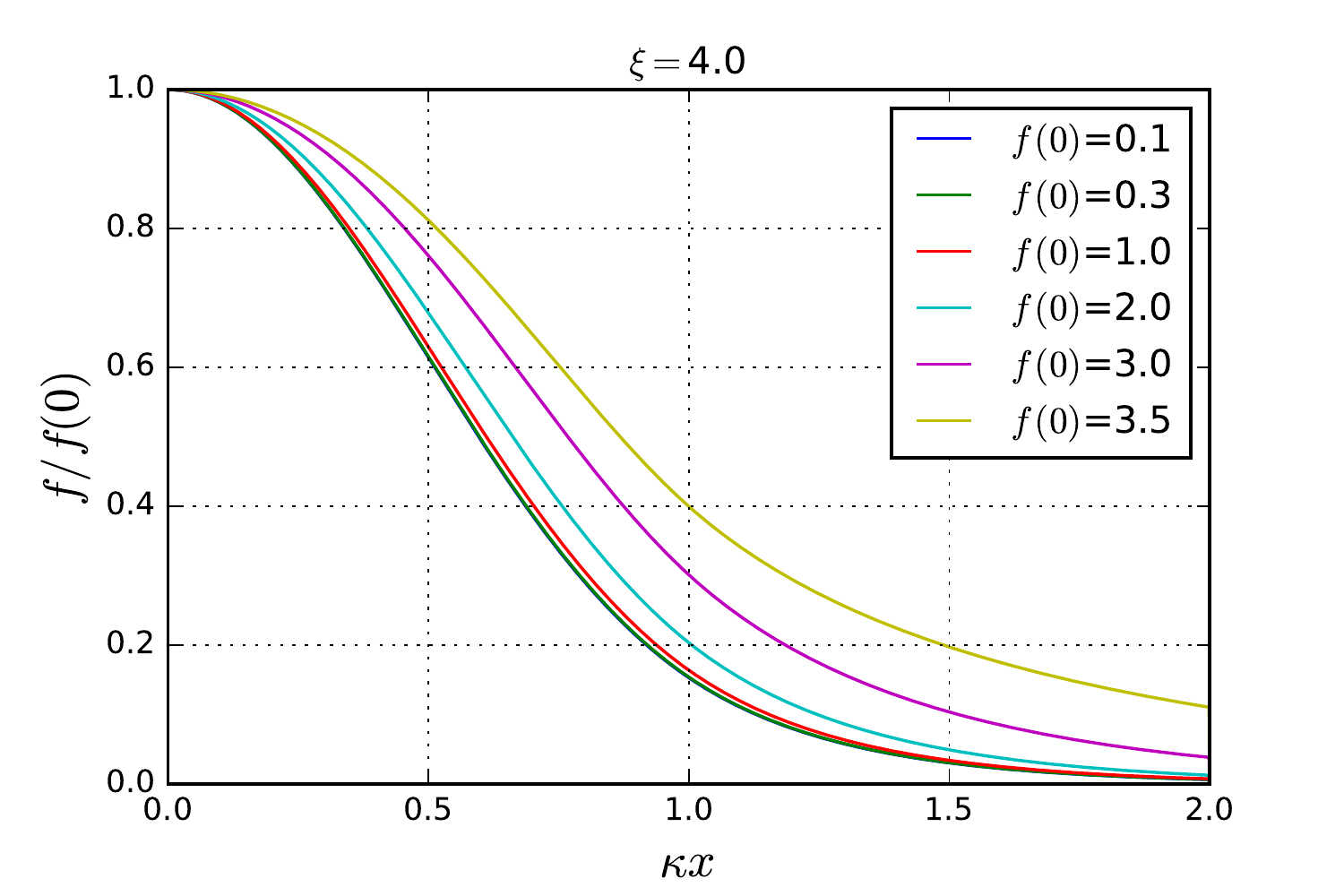}
\includegraphics[width=\fz]{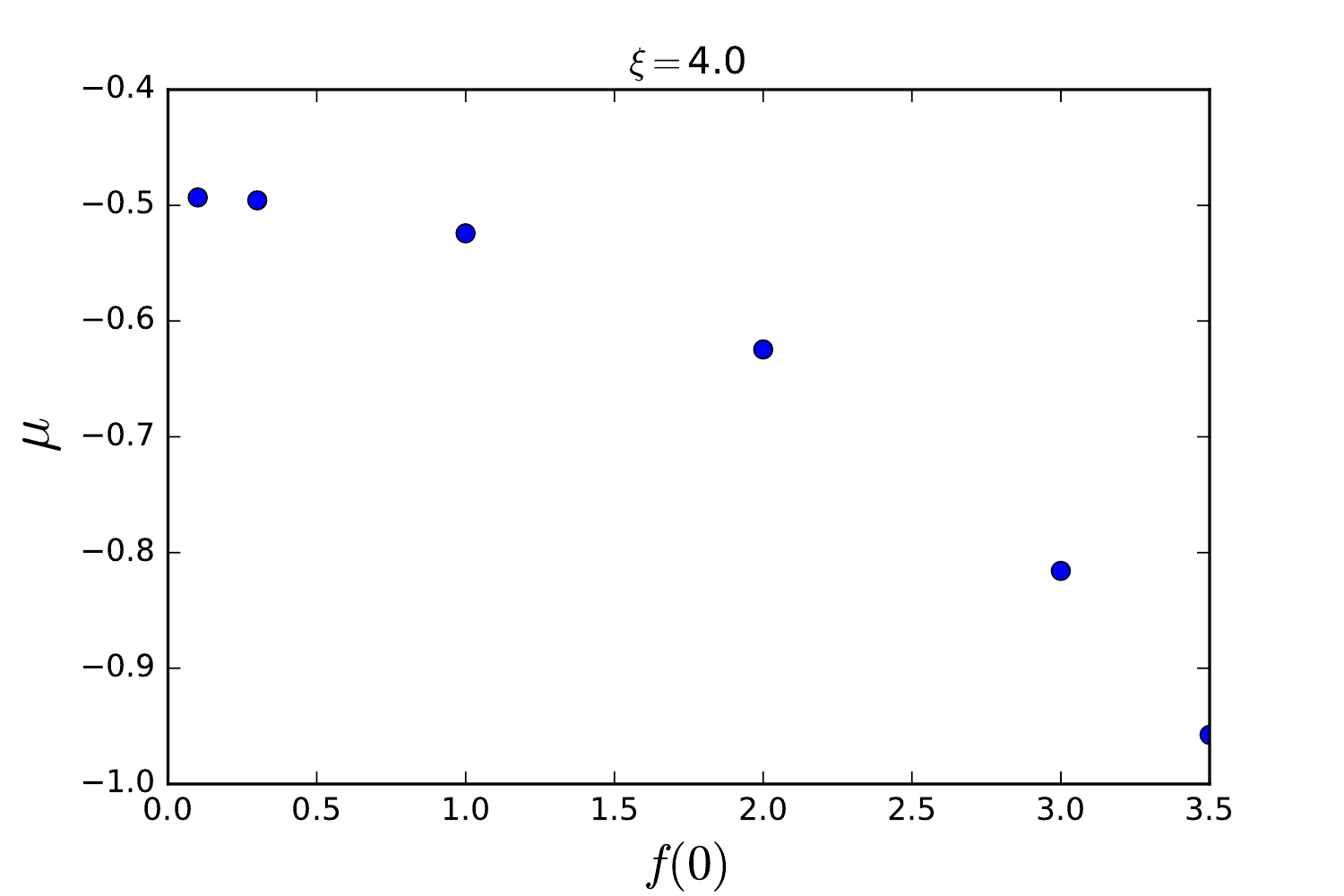}
\caption{Top panel: Fundamental solutions for $\xi = 4$ and different
values of $f(0)$, expressed in units of $\kappa / \sqrt{\gamma_2}$.
Bottom panel: Values of the parameter $\mu$ for these solutions. }
\label{fig:fun_NL}
\end{figure}

We solved Eq.~\eqref{eq:matching} numerically for various values of
$\xi$ and found few deviations from the above picture. In particular,
solutions with $m$ nodes exist for $m$ between $0$ and a maximum value
$m_\mathrm{max}$, approximately equal to $\xi / 4$ when $\xi \gg 1$.
We also solved Eq.~\eqref{eq:f2} numerically using a shooting method
to see the effects of the nonlinear term as well as that of a more
realistic profile for $h$. Concerning the former, we found its main
effect is to decrease the value of $\mu$ of each solution, by a term
quadratic in $f(0)$. For each value of $m$, there is a critical value
of $\left\lvert f(0) \right\rvert$ above which the solution disappears
as the corresponding value of $\mu$ drops below $-1$, as shown in
Fig.~\ref{fig:fun_NL} for the fundamental solution with $\xi = 4$. The
nonlinear term also has the tendency to widen the condensate, although
this becomes significant only close to the critical value.
Similarly, we found that replacing the above profile of $h$ with a
hyperbolic tangent does not change the qualitative behavior of the
solutions. Its main effect is to increase $m_\mathrm{max}$, which
seems to come from the slower convergence of $h$ towards $1$.

\subsection{Numerical construction}
\label{StudyNum}

We have solved numerically the coupled set of differential equations
\eqref{gauge}, \eqref{higgs} and \eqref{condensate}, subject to the
appropriate boundary conditions \eqref{Bound1} and \eqref{Bound2}. 

\begin{figure}
\includegraphics[width=\fz]{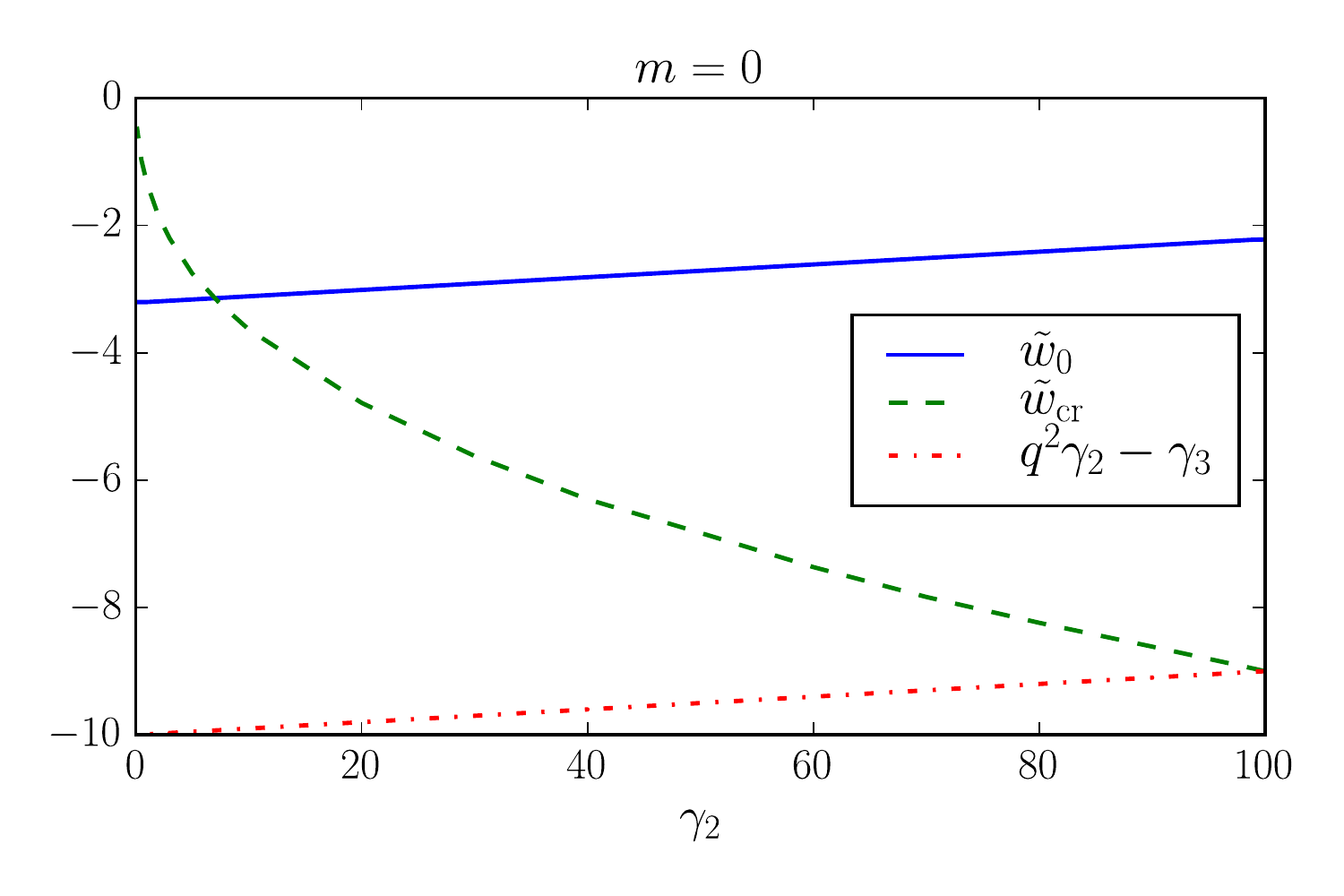}
\includegraphics[width=\fz]{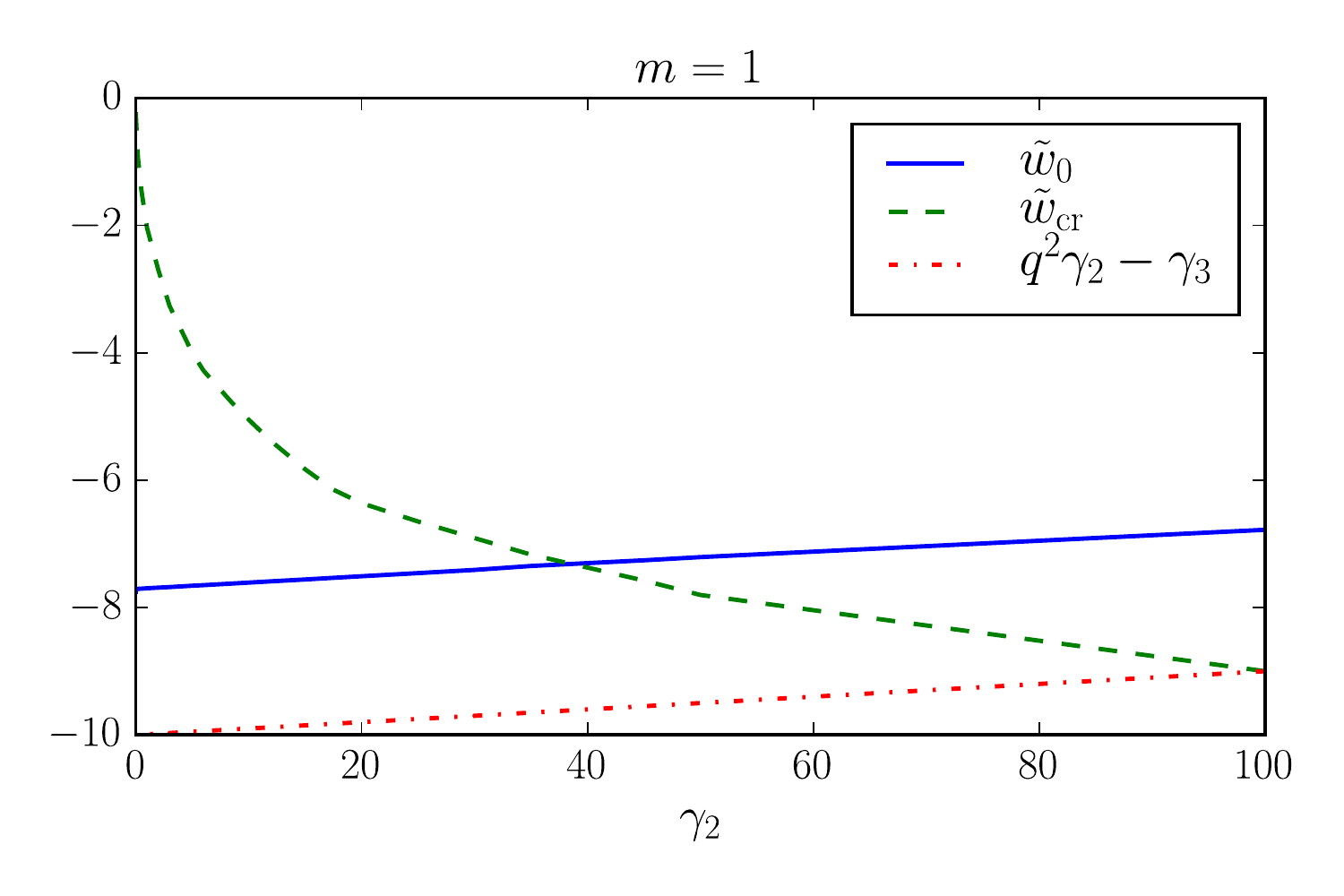}
\includegraphics[width=\fz]{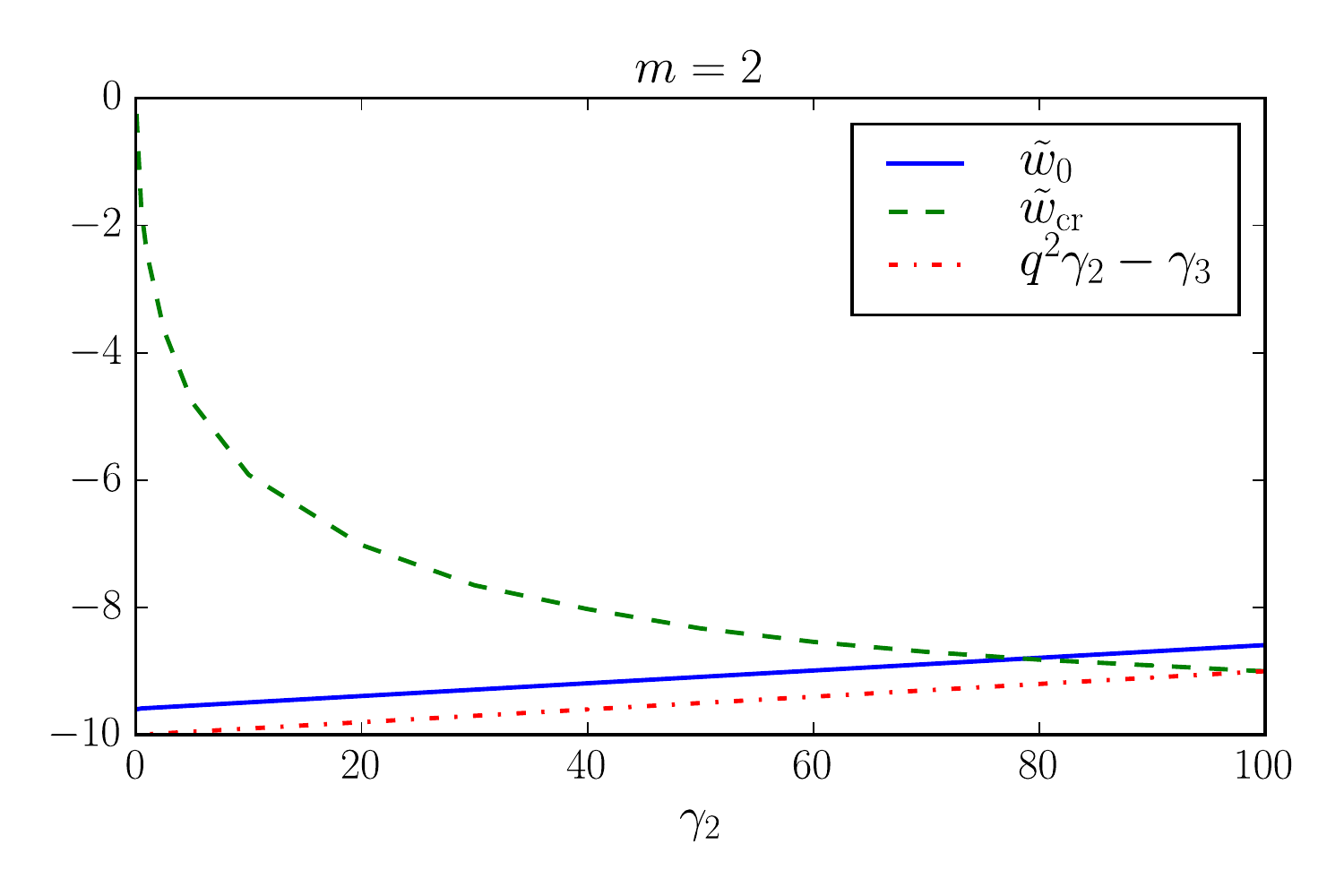}
\includegraphics[width=\fz]{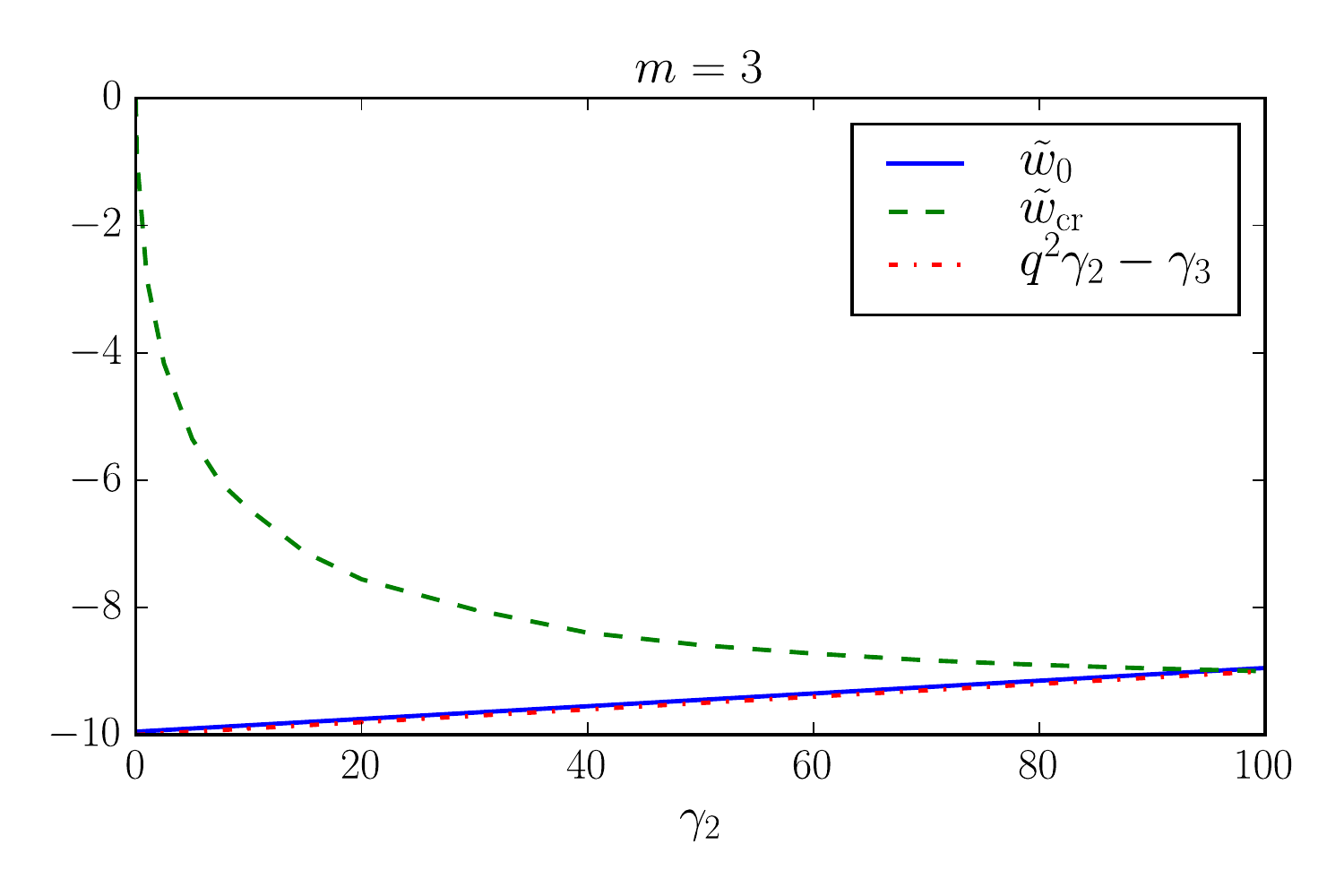}
\caption{The value of $\tilde{w}_\mathrm{cr}$ and $\tilde{w}_0$ in
dependence on $\gamma_2$ for $q=0.1$, $\gamma_3=10$, $\alpha_1=0.01$ and
$m=0,1,2,3$ (top to bottom), respectively. We also give the negative
value of the effective mass of the condensate function,
$-m_{\sigma}^2=q^2 \gamma_2 - \gamma_3$.}
\label{fig:om_0_om_cr}
\end{figure}

\subsubsection{A case study}

In what follows, we concentrate on the solutions for $q=0.1$,
$\alpha_1=0.01$ and $\gamma_3=10$ and study the effects of the variation
of $\gamma_2$. This case is complementary to the study done in
\cite{Hartmann:2016axn}, where the couplings $\gamma_2$ and $\gamma_3$
had been chosen one to two orders of magnitude larger. First let us
recall that restrictions on the couplings exist in this model, in
particular we have $\gamma_2 \leq \gamma_3^2$, such that in the
following we will study solutions for $\gamma_2 \in ]0:100[$. Note that
the second requirement $q^4 \gamma_2^2 \leq \gamma_2$  is automatically
fulfilled within this interval of the parameter $\gamma_2$.

We have  constructed solutions with up to 3 nodes in the condensate
field function.  We observe that for all values of $m$ solutions exist
in a limited interval of the central value of the condensate field,
$f(0) \in [0:f(0)_{\mathrm {max}}]$ such that for $f(0)\rightarrow 0$
the field function $f(x)\equiv 0$. This corresponds to a value of the
state parameter $\tilde{w}$ which we will denote $\tilde{w}_{0,m}$ in
the following. Our results for $m=0,1,2,3$ are shown in
Fig.~\ref{fig:om_0_om_cr}. In all plots, we show the (negative) value of
the effective mass of the condensate field, which is given by
$m_{\sigma}^2 = \gamma_3 - q^2 \gamma_2$. We observe that
$\tilde{w}_{0,m}$ is a linear function of $\gamma_2$ and is parallel to
$-\tilde{m}_{\sigma}^2$ for all values of $m$. The difference
$\Delta_{m} := \tilde{w}_{0,m} - (-\tilde{m}_{\sigma}^2)$ decreases with
increasing node number $m$. The values are given in Table \ref{table1}. 
For the given parameter values, we hence find that the 
formula
\begin{equation*}
\tilde{w}_{0,m}=\Delta_m - \gamma_3 + q^2 \gamma_2 
\end{equation*}
holds. 

Our numerical results indicate that $f(0)$ can be increased up to a
maximal value $f(0)_{\mathrm{max}}$ which depends on the values of the
couplings in the model. We will denote the corresponding value of the
state parameter $\tilde{w}_{\mathrm{cr},m}$ in the following. The value
of $\tilde{w}_{\mathrm{cr},m}$ is a decreasing function of $\gamma_2$.
The {\it qualitative} behaviour is similar for all values of $m$~:
$\tilde{w}_{\mathrm{cr},m}=0$ for $\gamma_2\rightarrow 0$ and decreases
to $\tilde{w}_{\mathrm{cr},m}=-9$ for $\gamma_2\rightarrow 100$, where
it meets with the curve for $-m_{\sigma}^2$.

Let us denote the value of $\gamma_2$ at which
$\tilde{w}_{\mathrm{cr},m}=\tilde{w}_{0,m}$ by
$\gamma^{(\mathrm{eq},m)}_{2}$, the numerical values of which are given
in Table \ref{table1} for $m=0,1,2,3$. 

\begin{table}
\begin{tabular}{| c | c | c | }
  \hline			
  $m$ & $\Delta_m$ & $\gamma_2^{(\mathrm{eq},m)}$\\
  \hline
  $0$ & $6.70$  & $7$  \\
  $1$ & $2.20$  & $34$ \\
  $2$ & $0.41$ & $ 80$ \\
  $3$ & $0.05$ & $94$ \\
  \hline
\end{tabular}
\label{table1}
\caption{Some characteristic values of the $\tilde{w}_{0,m}$ and
$\tilde{w}_{\mathrm{cr},m}$ curves shown in Fig.~\ref{fig:om_0_om_cr}. } 
\end{table}

\begin{figure}
\includegraphics[width=\fz]{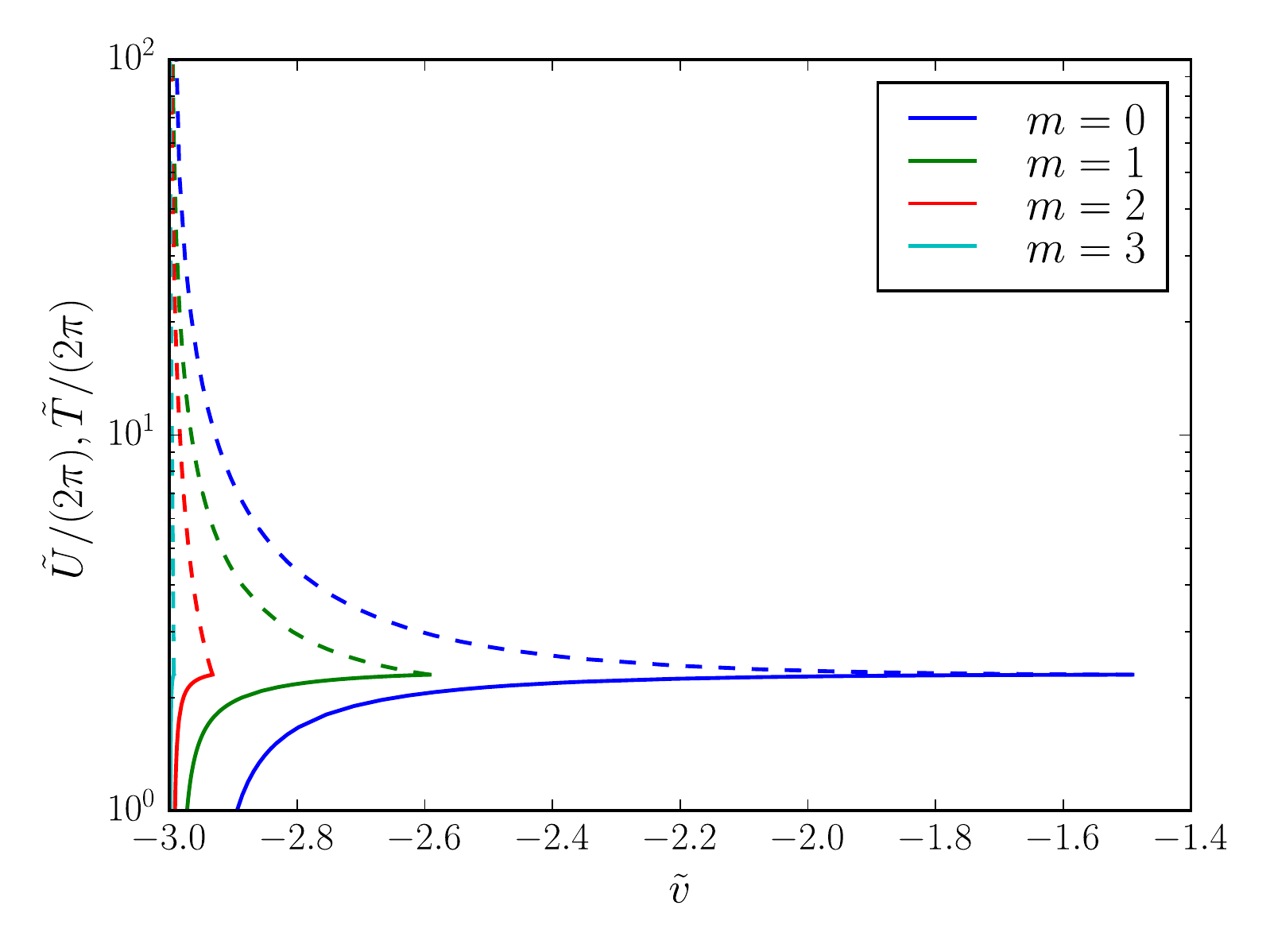}
\includegraphics[width=\fz]{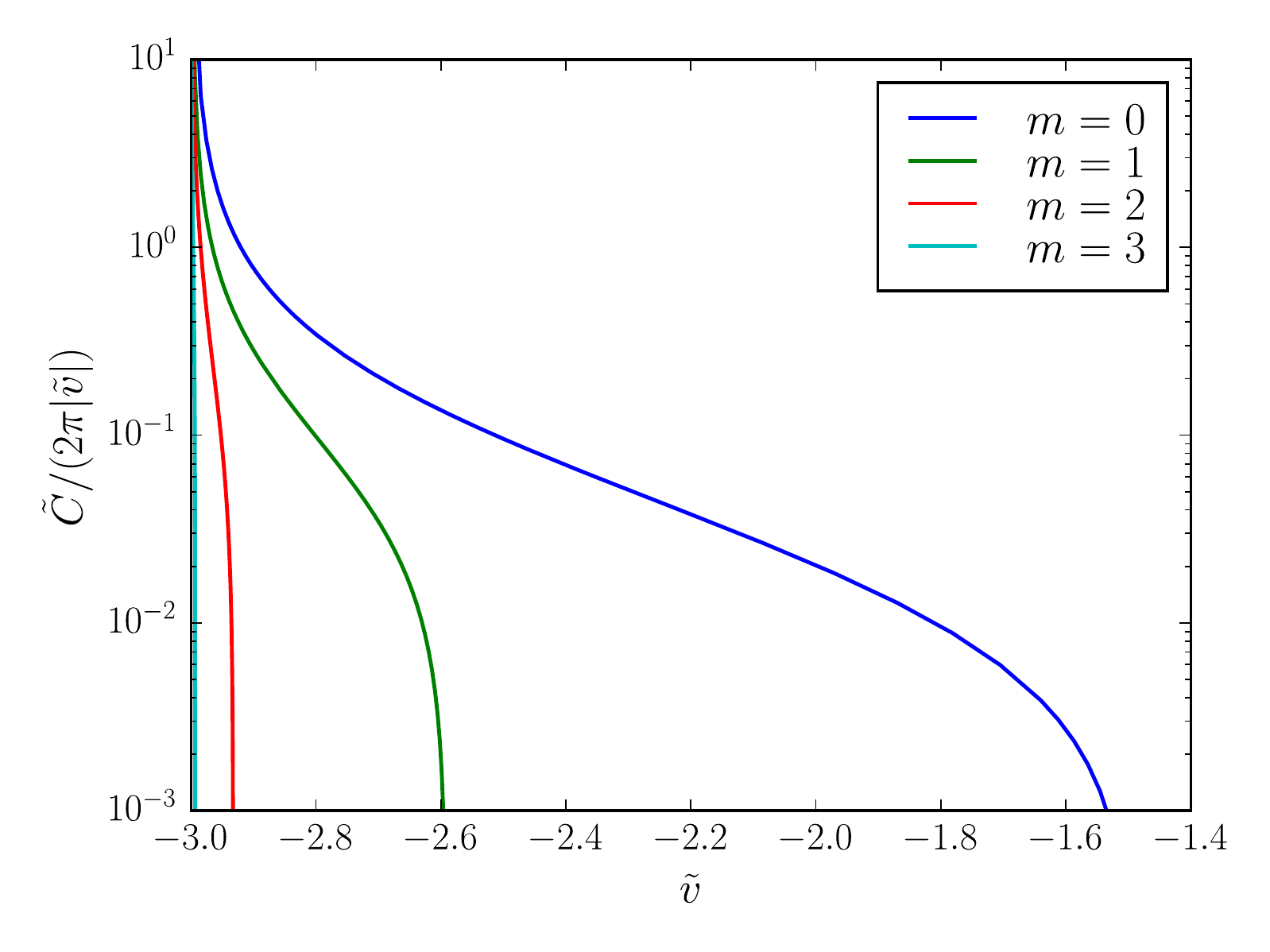}
\caption{Top~: The energy per unit length $\tilde{U}$ (dashed) and the
tension $\tilde{T}$ (solid) as function of the state parameter
$\tilde{v}$ for $\alpha_1=0.01$, $q=0.1$, $\gamma_2=99$, $\gamma_3=10$
and $m=0, 1, 2, 3$. Bottom~: Same as top for the current $\tilde{C}$. The
phase frequency threshold of Eq.~\eqref{OmegaBound} is clearly visible
as the divergence point at $\tilde{w}_\mathrm{th}=-\gamma_3 + q^2 \gamma_2
\simeq -3$.}
\label{fig:UTC}
\end{figure}

For $\gamma_2=\gamma_2^{(\mathrm{eq},m)}$, the qualitative dependence of
$\tilde{w}$ on the central condensate value $f(0)$ changes. For
$\gamma_2 > \gamma_2^{(\mathrm{eq},m)}$ the state parameter $\tilde{w}$
decreases for increasing $f(0)$ such that for $\tilde{w}\rightarrow
\tilde{w}_{\mathrm{cr},m}$ the value of the condensate $f(0)$ becomes
very large and, in fact, as our numerical results indicate, tends to
infinity, $f(0)_\mathrm{max}\rightarrow \infty$. This case has been
studied in detail in \cite{Hartmann:2016axn}. Here we present our
results for the energy per unit length $\tilde{U}$, the tension
$\tilde{T}$ and the current $\tilde{C}$ as functions of the state
parameter $\tilde{v}$ for $\alpha=0.01$, $q=0.1$, $\gamma_2=99$ and
$\gamma_3=10$ in Fig.\ref{fig:UTC}.

For increasing $m$ the range in $\tilde{v}$, for which superconducting
string solutions exist decreases. At $\tilde{v}_{\rm 0}$ the energy per 
unit length and tension are equal and the current $\tilde{C}$ becomes
zero. At $\tilde{v}_{\rm cr}$ the current diverges. We find that
independent of the value of $m$, $\tilde{v}_{\rm cr} = -3$ and that the
maximal value of $f(0)_{\rm max}$ corresponding to this critical value
is (nearly) independent of the node number. Given the interpretation put
forward in \cite{Carter:1994xr}, namely considering the current
$\tilde{C}$ and $\tilde{v}$ as a conjugate pair, in which $|\tilde{v}|$
is the particle number density and $\tilde{C}$ the chemical potential or
effective mass per particle, we find that solutions exist only above a
certain particle number density, which increases with increasing node
number $m$. Furthermore, for a given particle number density
$|\tilde{v}|$, the effective mass per particle is largest for the $m=0$
solution and decreases with increasing node number. At the maximal
possible particle number $|\tilde{v}_{\rm cr}|$ the current diverges.

 For $\gamma_2 <
\gamma_2^{(\mathrm{eq},m)}$, on the other hand, we find that $f(0)$ can
be increased up to a maximal value $f(0)_\mathrm{max}$, which is
finite, and that $\tilde{w}$ is an increasing function of $f(0)$. From
this maximal value of $f(0)$ a second branch of solutions exists for
decreasing $f(0)$, while the state parameter $\tilde{w}$ further
increases. We will discuss the origin of the existence of this branch
and the physical phenomena associated to it in subsection
\ref{sub:2branch}.

\subsubsection{Higgs field oscillations}

\begin{figure}
\includegraphics[width=\fz]{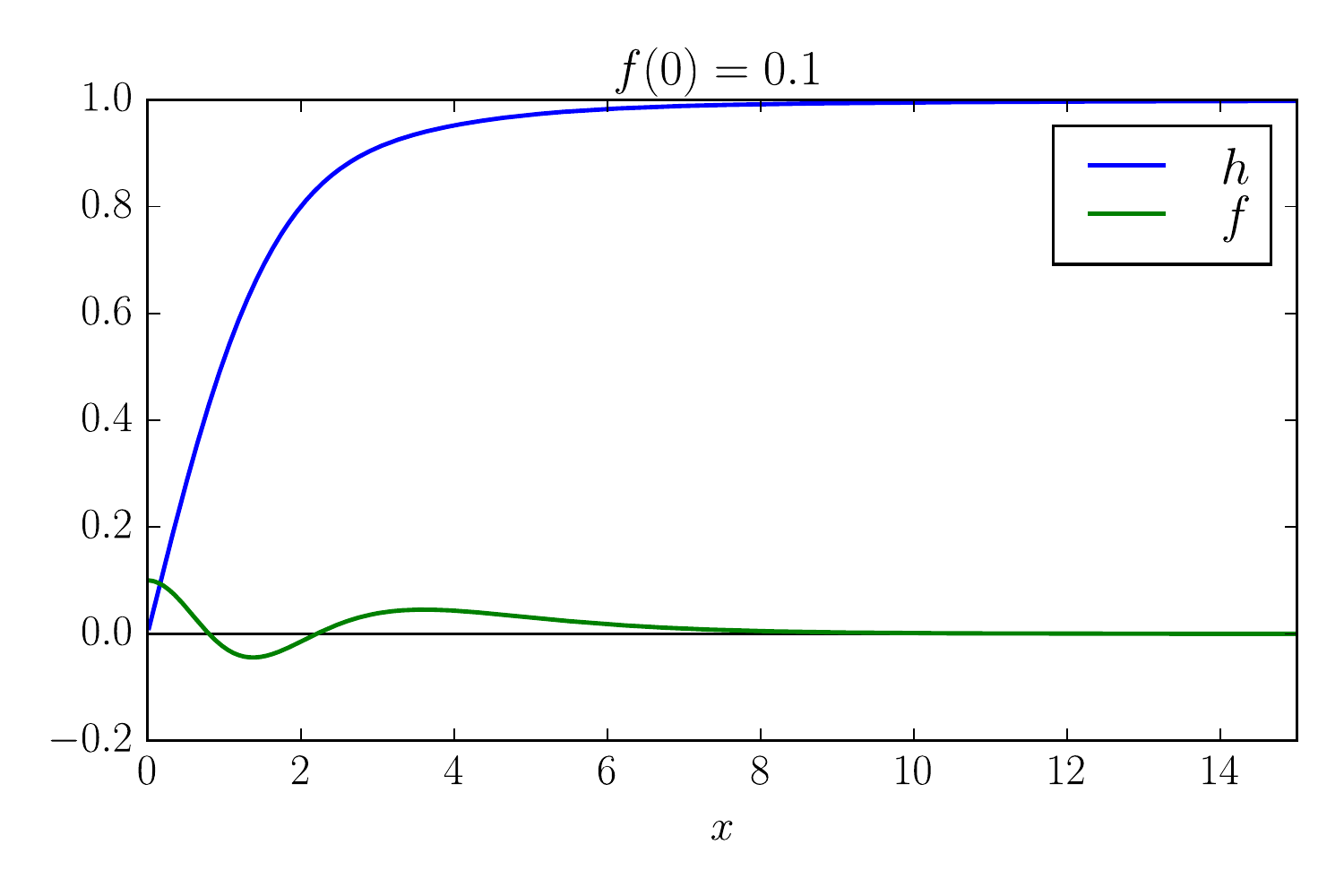}
\includegraphics[width=\fz]{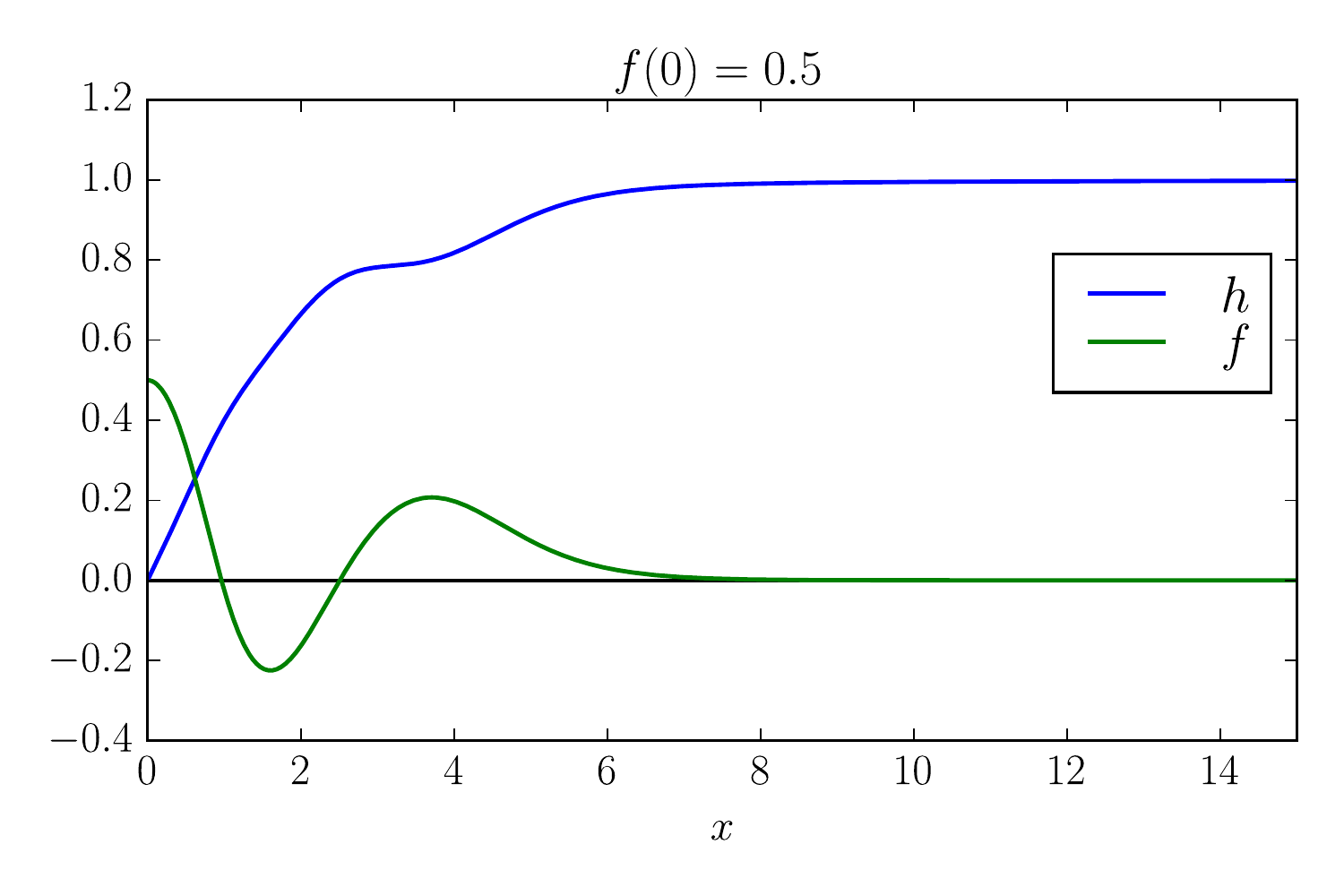}
\includegraphics[width=\fz]{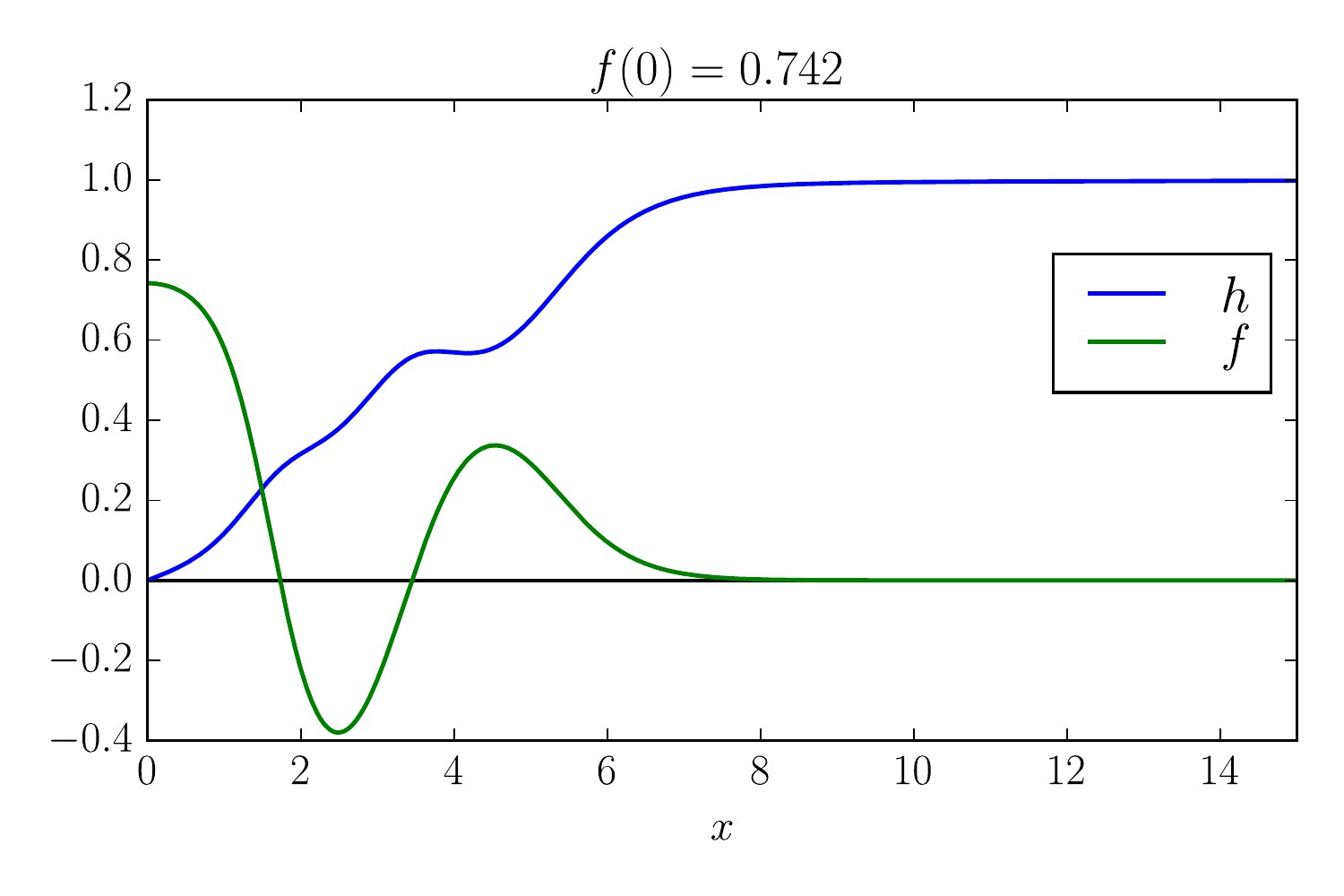}
\caption{The string-forming Higgs field profile $h(x)$ and the condensate
$f(x)$ as functions of the rescaled core radius $x$ for 
$\alpha_1=0.01$, $q=0.1$, $\gamma_2=\gamma_3=10$, $m=2$ and various
values of $f(0)\in \{ 0.1,0.5,0.742\}$ (from top to bottom).}
\label{fig:m2_higgs}
\end{figure}

During the study of smaller values of the couplings $\gamma_2$ and
$\gamma_3$, we observed a new phenomenon that is not present for the
cases presented in \cite{Hartmann:2016axn}. The reason for this is that
the central value of the condensate function, $f(0)$ can have larger
values for smaller values of $\gamma_2$ and $\gamma_3$, respectively.
For sufficiently large values of $f(0)$ we find that the oscillations of
the condensate function can trigger an oscillatory behavior in the Higgs
field function. This is shown for $\gamma_2=\gamma_3=10$, $q=0.1$,
$\alpha_1=0.01$ and $m=2$ in Fig.~\ref{fig:m2_higgs}, in which we also
show the condensate field function $f(x)$ together with the Higgs field
function $h(x)$ for increasing values of $f(0)$ up to the maximal
possible value of $f(0)\approx 0.742$. As can be clearly seen here, the
Higgs field function increases monotonically from zero at the origin to
unity at infinity for small values of $f(0)$, here $f(0)=0.01$ and
$f(0)=0.1$. But as soon as $f(0)$ is large enough, we see that the Higgs
field starts to show an oscillating behaviour, see the profiles for
$f(0)=0.5$ and $f(0)=0.742$.

\begin{figure}[h]
\centering
\includegraphics[width = \fz]{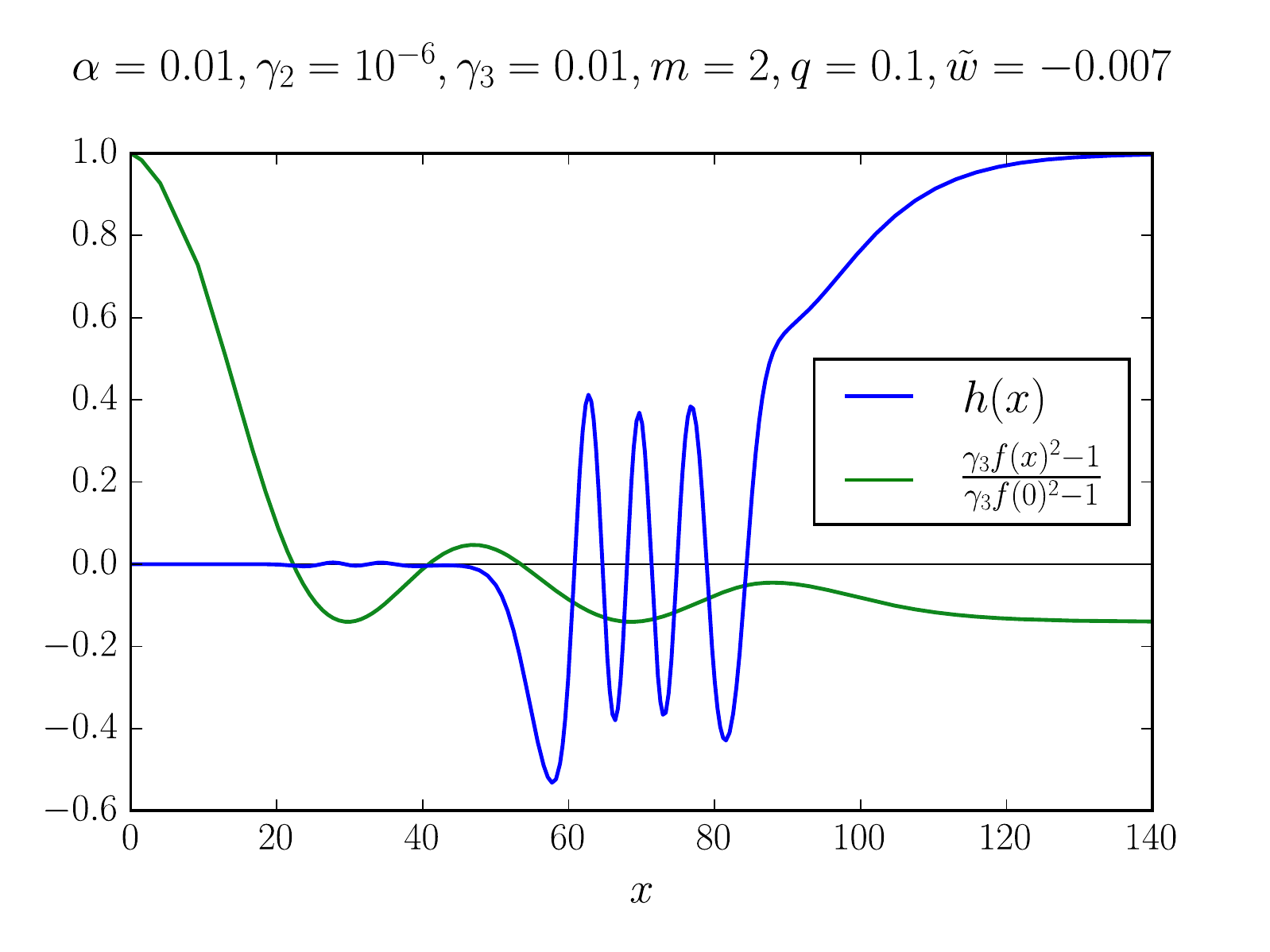}
\caption{The Higgs field function $h(x)$ together with $\lc\gamma_3
f^2(x)-1\rc/\lc\gamma_3 f^2(0) -1\rc$ for $m=2$, $\alpha_1=0.01$,
$\gamma_2=10^{-6}$, $\gamma_3=0.01$, $q=0.1$ and $\tilde{w}=-0.007$.   }
\label{fig:higgs_oscillation_m2}
\end{figure}

For values of $\gamma_2$ and $\gamma_3$ even smaller -- and consequently
$f(0)$ much larger -- we observe oscillations of the Higgs field with
large relative amplitudes on a finite interval of the radial coordinate
$x$, on which the Higgs field function possesses nodes. As a first
approximation this can be understood by considering (\ref{higgs}) and
assuming that the terms in $P$ and $h^3$ can be neglected with respect
to the $f^2$ term.  Assuming further that the oscillations appear away
from the origin $x=0$, we can also neglect (to first approximation) the
$h'$ term such that the equation reads $h''\approx h\lc \gamma_3 f^2(x)
-1\rc$, which has oscillating solutions for $\gamma_3 f^2(x) -1 < 0$. Our
numerics  confirms this and we find that the Higgs field oscillations
occur in an interval of $x$ which is bounded by those two values of $x$
for which $\gamma_3f^2(x) - 1 =0$. This is shown in
Fig.~\ref{fig:higgs_oscillation_m2} for $m=2$, $\alpha_1=0.01$
$\gamma_2=10^{-6}$, $\gamma_3=0.01$, $q=0.1$ and $\tilde{w}=-0.007$,
i.e. a solution very close to the chiral limit. This solution has
central condensate value $f(0)\approx 28.5$ and clearly possesses
oscillations of the Higgs field  in the two intervals of $x$, where
$\gamma_3 f^2(x)-1 < 0$.

We have also investigated cases with different values of $m  > 0$ and
$\gamma_3$ and confirm that for the parameter range studied here, the
back-reaction of the condensate function induces oscillations in the
Higgs field function when $f(0)$ becomes large, i.e. when non-linear
back-reaction effects can no longer be neglected.

Note that we do {\it not} observe oscillations in the limit
$f(0)\rightarrow 0$ and/or for $m=0$, hence this phenomenon is
restricted to a regime of the parameter space which allows for large
values of $f(0)$. We believe this phenomeon to be very rich and to have
important implications.  A detailed numerical analysis, which is outside
the scope of this paper, is hence left as future work.

\subsubsection{The second branch}
\label{sub:2branch}

\begin{figure}[h]
\centering
\includegraphics[width = \fz]{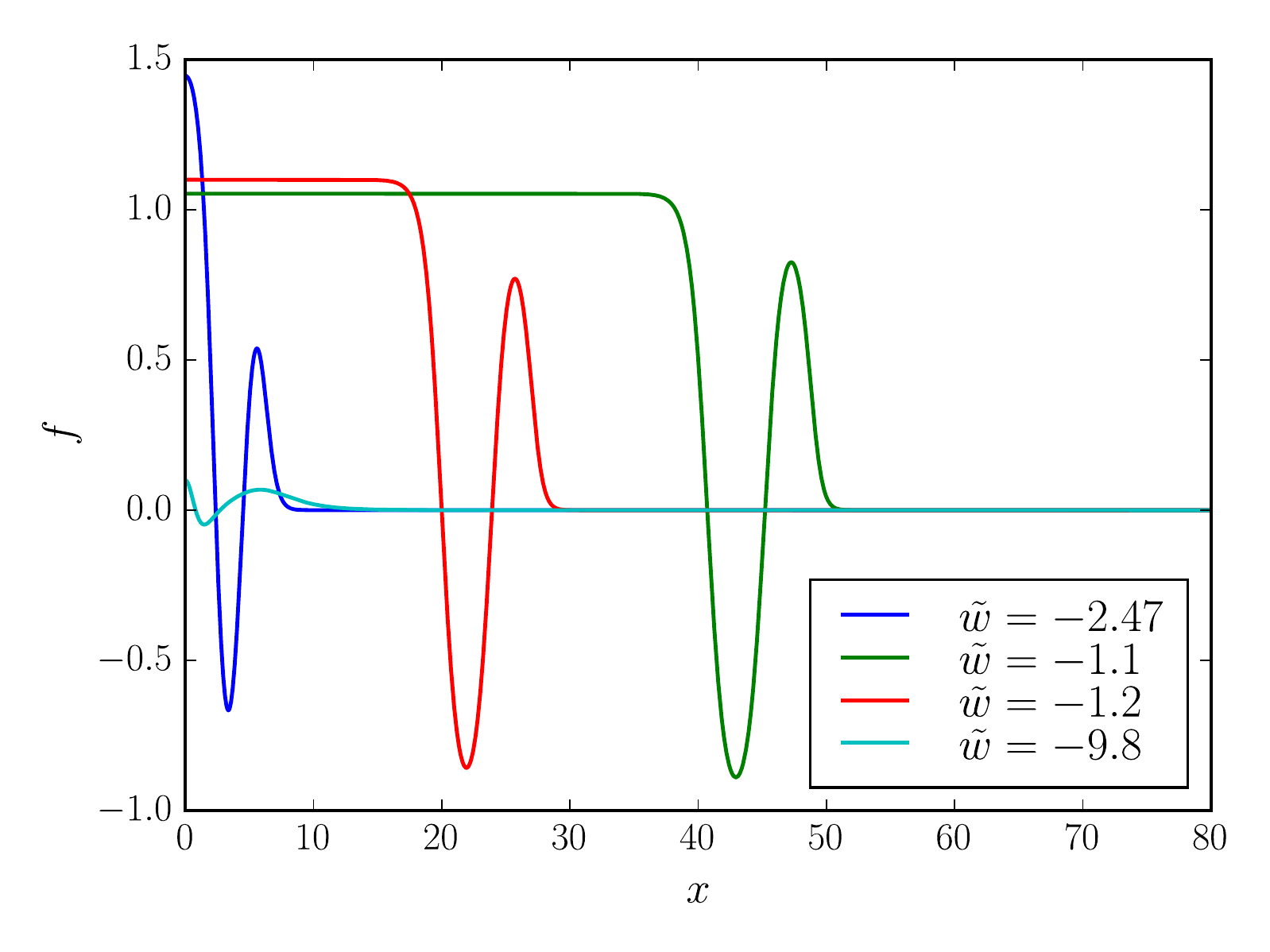}
\caption{The condensate field function $f(x)$ for $m=2$, $\alpha_1=1.0$,
$\gamma_2=1.0$, $\gamma_3=10.0$, $q=0.1$ for increasing values of $\tilde{w}$. 
Note that $\tilde{w}=-2.47$ corresponds to $\tilde{w}_{\mathrm{cr}}$.   }
\label{fig:condensate_m2}
\end{figure}

When increasing the central value $f(0)$ of the condensate function we
observe that the structure associated to the oscillation of the
condensate function remains close to the string axis. This changes on
the second branch of solutions mentioned above. Decreasing $f(0)$ from
its maximal value, the value of $\tilde{w}$ increases further on the
second branch. We observe that, although the value of $f(0)$ decreases,
it does so slowly. However, with increasing $\tilde{w}$ the structure
associated to the condensate field oscillations moves to larger values
of $x$, i.e. we obtain solutions with $h(x)\approx 0$ and $f(x) \approx
\mathrm{constant} \lesssim  f(0)_{\mathrm{max}}$ on an interval $x\in
[0:\delta]$, where $\delta$ increases with increasing $\tilde{w}$.  This
is shown for $m=2$, $\alpha_1=1.0$, $\gamma_2=1.0$, $\gamma_3=10$,
$q=0.1$ and increasing value of $\tilde{w}$ in
Fig.~\ref{fig:condensate_m2}. For $w=-9.8$  the value of $f(0)=0.1$.
Increasing $\tilde{w}$ up to $\tilde{w}_{\mathrm cr}=-2.47$ the value of
$f(0)$ and with that the condensate and current increase close to the
string axis. The maximal possible value of $f(0)$ in this case is
$f(0)=f(0)_{\mathrm max}\approx 1.45$. Increasing $\tilde{w}$ further
leads now to the decrease of $f(0)$ and the increase of $\delta$. For
$\tilde{w}=-1.2$ and $\tilde{w}=-1.1$, respectively, we find $f(0)=1.1$
and $f(0)=1.05$.

\begin{figure}[h]
\centering
\includegraphics[width = \fz]{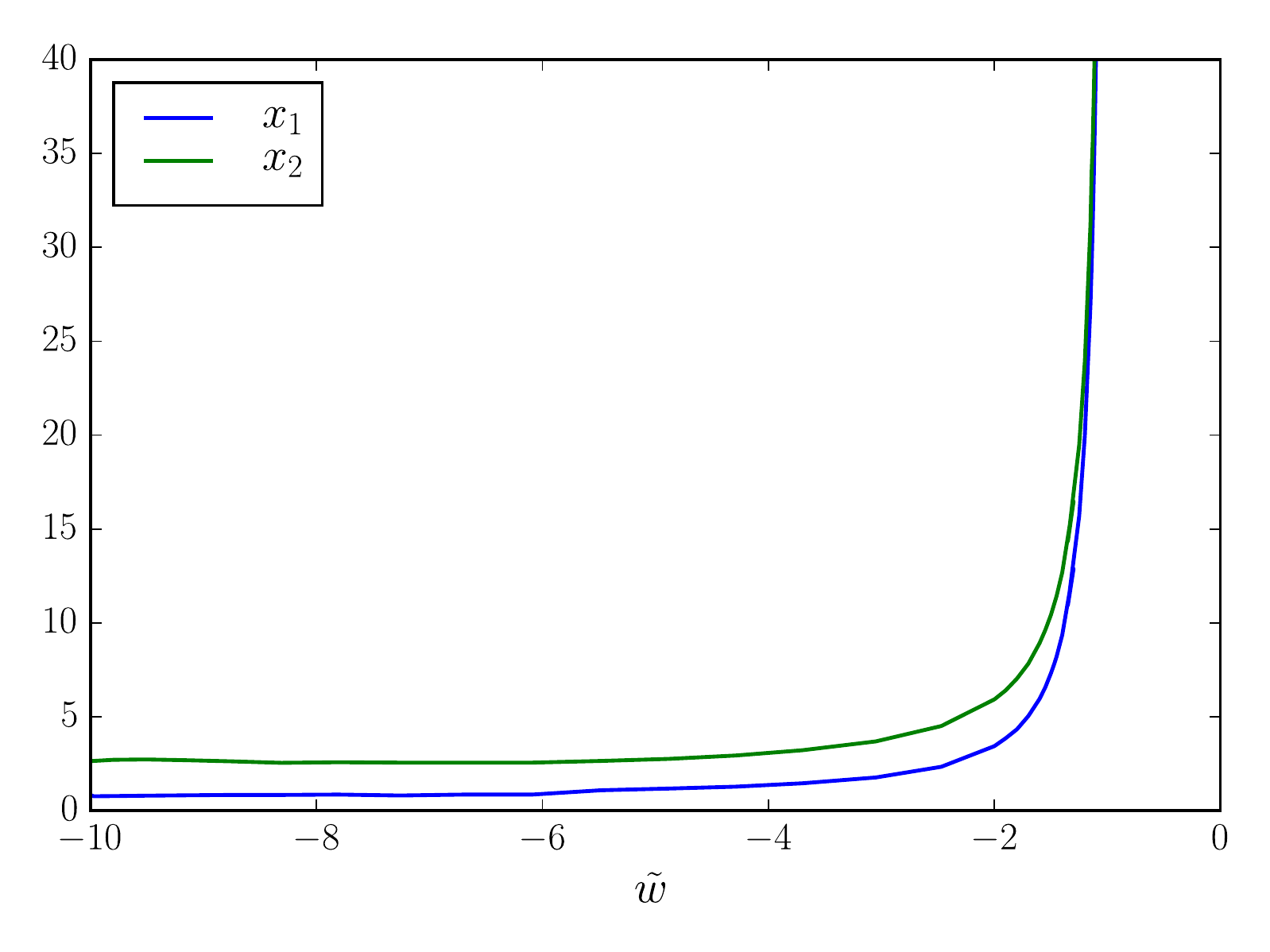}
\caption{The values $x_1$ and $x_2$  of the first and second node of the
condensate function $f(x)$ in dependence on $\tilde{w}$ for  $m=2$,
$\alpha_1=1.0$, $\gamma_2=1.0$, $\gamma_3=10.0$  and $q=0.1$.}
\label{fig:nodes}
\end{figure}

The fact that the structure moves out to infinity can also be clearly
seen when investigating the location of the zeros of the condensate
function. This is shown for a solution with $m=2$ nodes, $\alpha_1=1.0$, 
$\gamma_2=1.0$, $\gamma_3=10.0$, $q=0.1$ in Fig.~\ref{fig:nodes}, where
we give the positions $x_1$ and $x_2$, respectively, of the two nodes in
dependence of $\tilde{w}$.

\begin{figure}
\includegraphics[width=\fz]{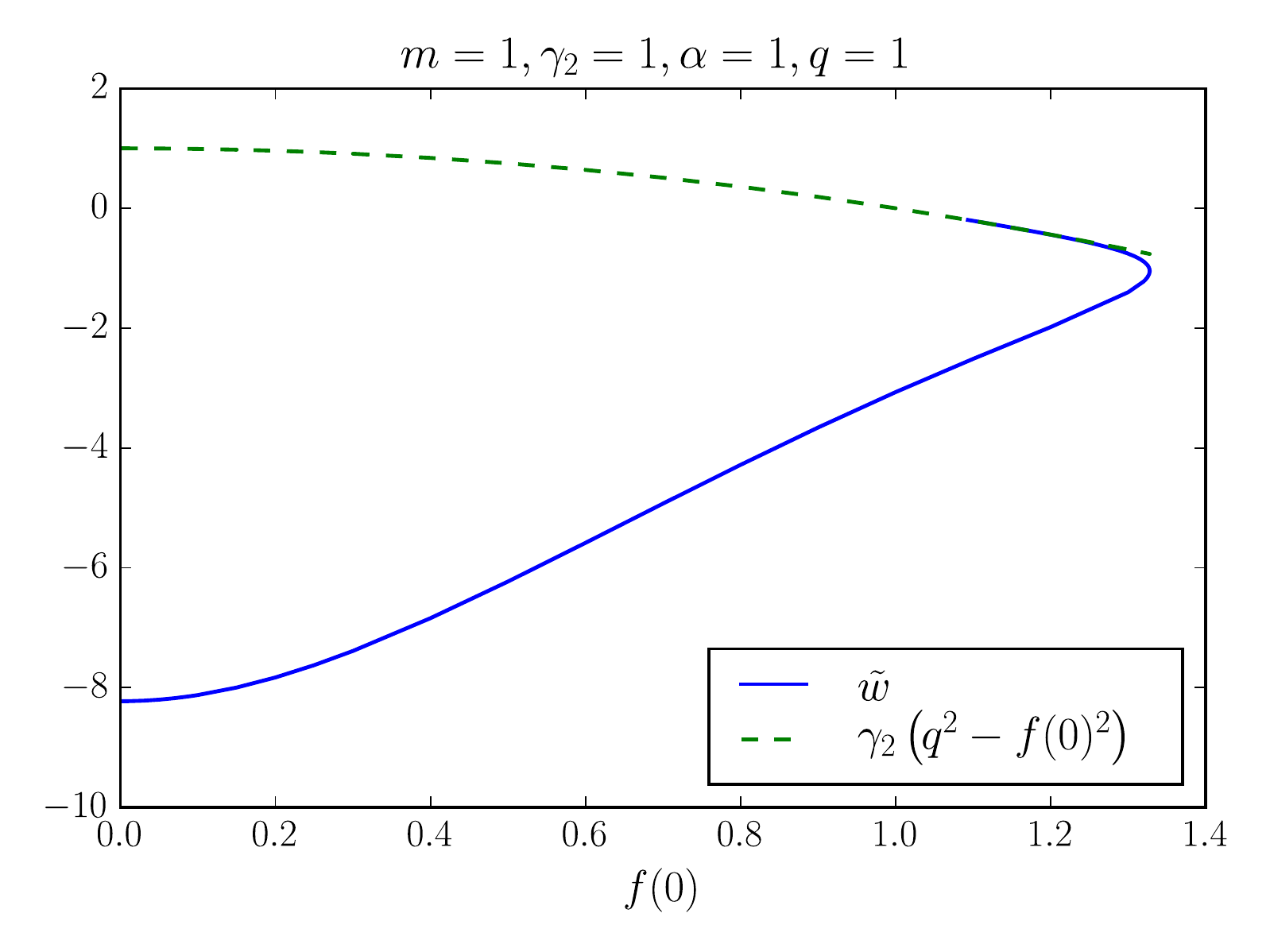}
\includegraphics[width=\fz]{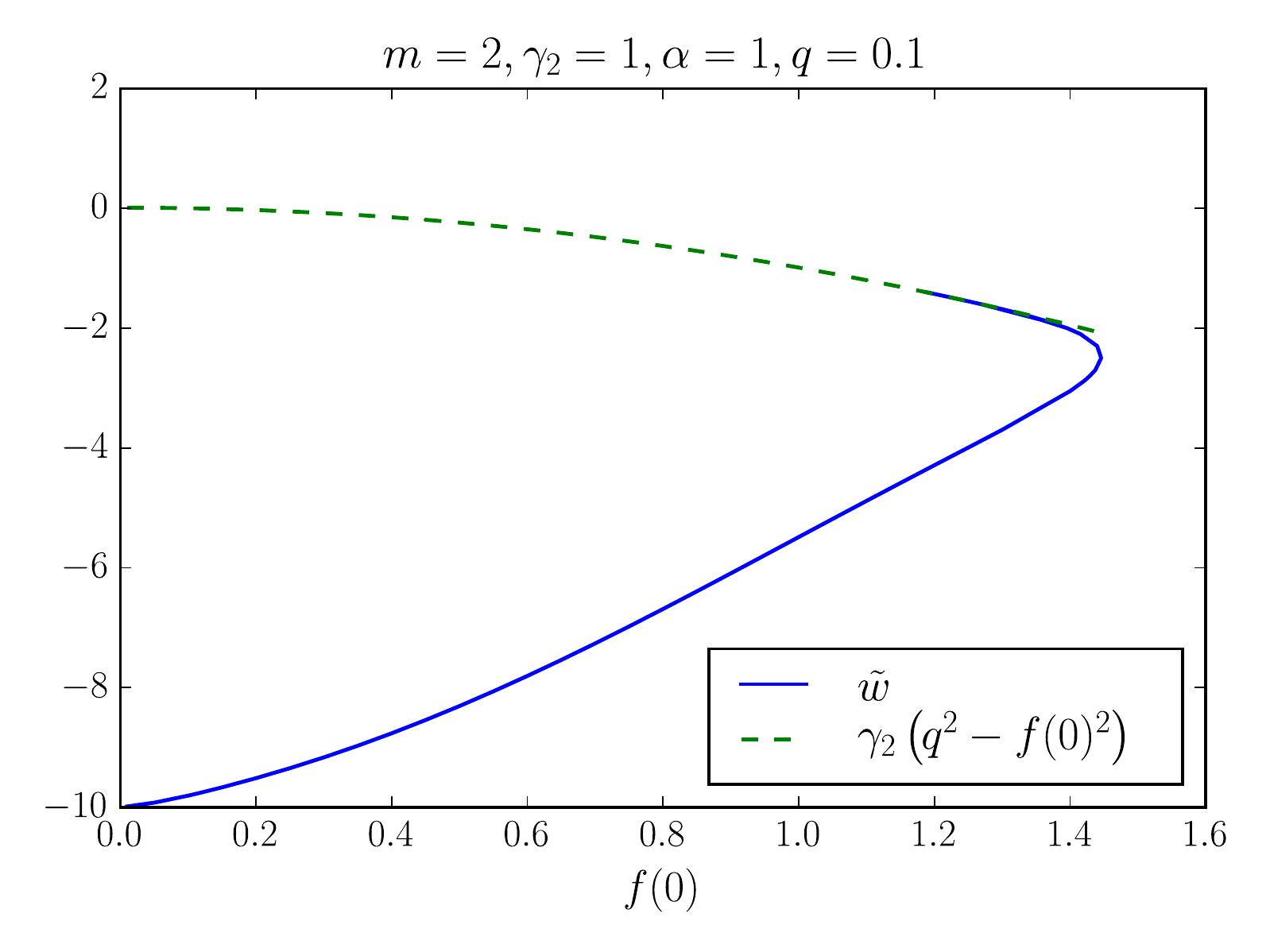}
\includegraphics[width=\fz]{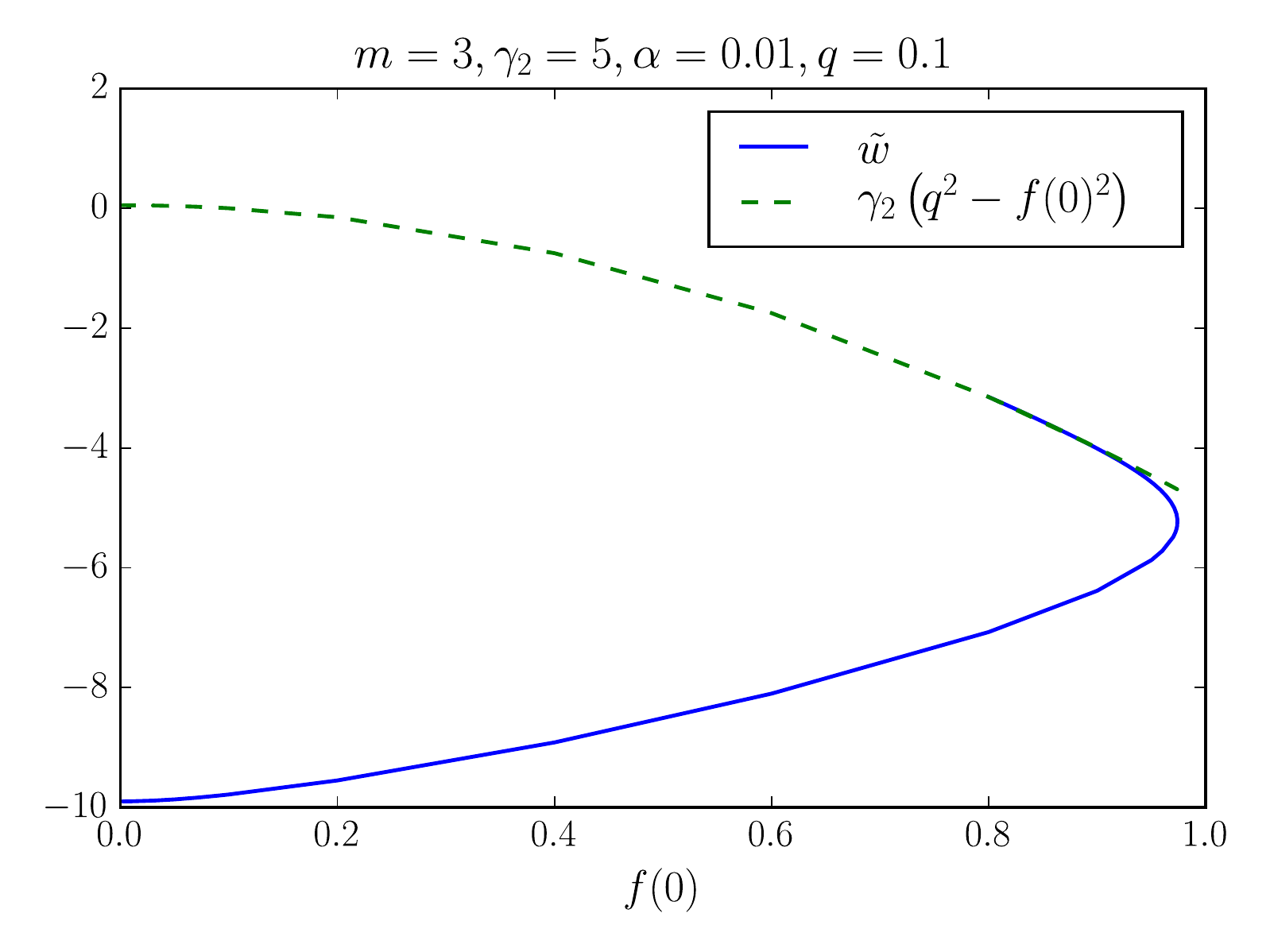}
\caption{The value of $\tilde{w}$ as function of $f(0)$ for three different
sets of parameter choices with $\gamma_3=10$. 
We give the numerical data (solid blue) as well as the analytic
curve $\gamma_2(q^2 - f(0)^2)$ (dashed green).   }
\label{fig:w_vs_f0}
\end{figure}

Decreasing $\tilde{w}$ on the second branch of solutions turns out to be
numerically very difficult, but we believe it to be very reasonable that
this second branch can be extended backwards all the way to $f(0)=0$, in
the limit of which the structures moves to infinity and the energy per
unit length $U$ and the tension $T$ tend to infinity. We can understand
this dependence by considering the condensate field equation
(\ref{condensate}) on the interval $x\in ]0:\delta]$, where $h(x)\equiv
0$ and $f(x)=\mathrm{constant.}=f(0)$ Excluding the possibility
$f(x)\equiv 0$, this implies that $\tilde{w} +  \gamma_2 \left[f(0)^2 -
q^2\right]=0$.  We demonstrate that our numerical data joins this curve
for three different sets of parameters, see Fig.~\ref{fig:w_vs_f0}.
Hence, though the numerics becomes very hard at the end points of the
respective numerical data curves, the analytically given curves are
(very likely) the proper continuation. We do not see any indications in
the numerics that the curves should stop.

Finally, let us explain {\it qualitatively} why two branches of
solutions in $f(0)$ exist in our model. This is easily understood when
remembering that we have rescaled the radial coordinate $r\rightarrow x
=r/\Lambda$ as well as the state parameter $w\rightarrow
\tilde{w}=\Lambda^2 w$, where $\Lambda=(\sqrt{\lambda_1} \eta_1)^{-1}$
is the length scale associated to the Higgs field. When increasing
$\tilde{w}$ on the first branch of solutions, we increase the condensate
close to the string axis until we reach the maximal possible value of
the condensate related to a value of $\tilde{w}$, which stays fixed on
the second branch of solutions and is negative in the case studied
above.  Now to increase the value of $\tilde{w}$ further, i.e. make it
tend to zero from below, we need to decrease $\Lambda$. But this in turn
implies that the rescaled radial  coordinate $x$ increases. This is
exactly what we observe in our numerics -- the non-trivial structure in 
the fields moves out to larger values of $x$.

\subsubsection{Strings with $n > 1$}
\label{sub:largern} 

We have also constructed superconducting string solutions with $n > 1$,
motivated by a recent study done in a very similar model
\cite{Forgacs:2016ndn,Forgacs:2016iva}. As our stability analysis below
shows, the qualitative behaviour of our results is independent of $n$.
To demonstrate this, we have constructed numerically solutions with
$n=2$ and $n=3$ and compared these to the $n=1$ case.

\begin{figure}
\includegraphics[width=\fz]{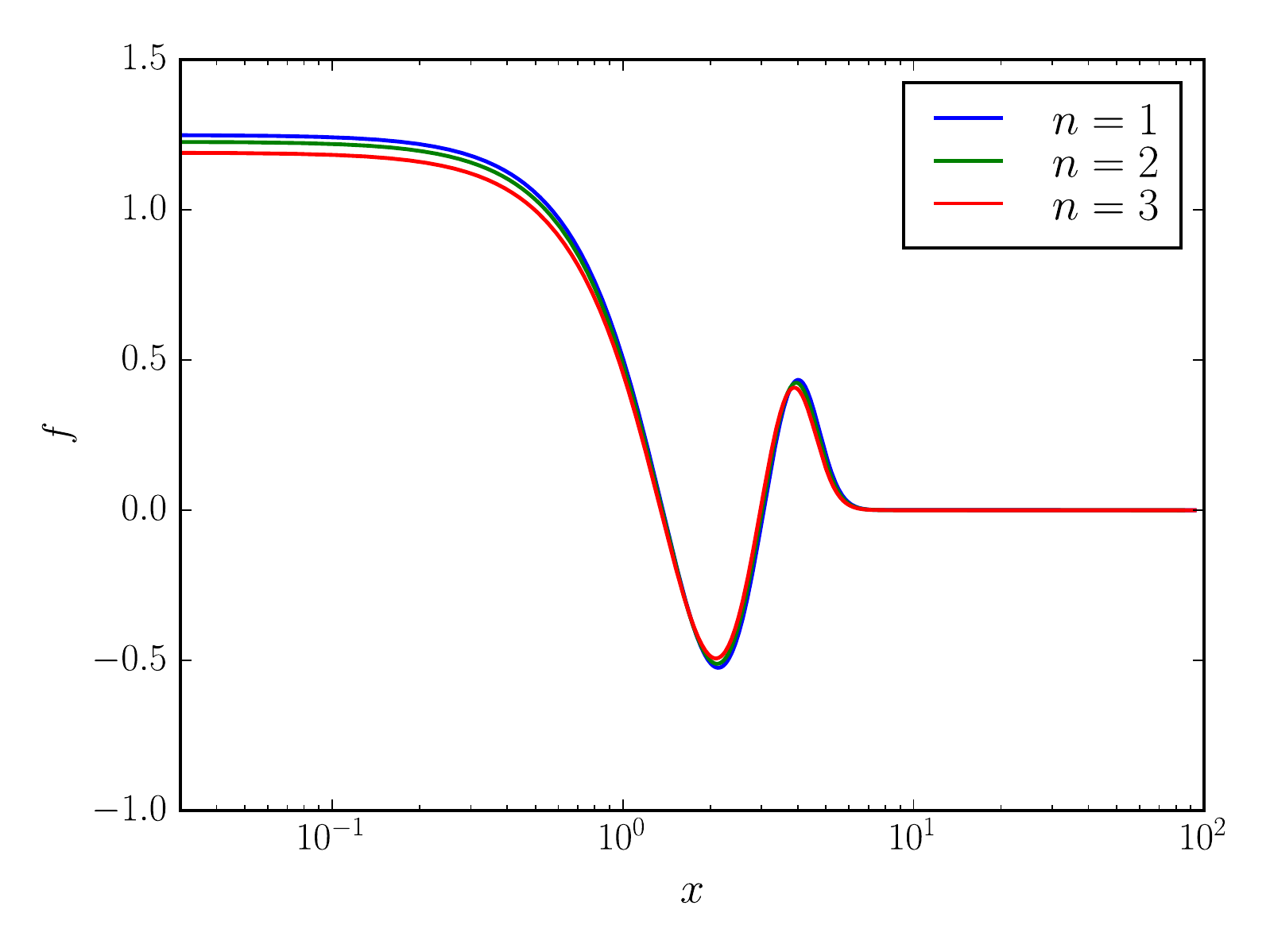}
\includegraphics[width=\fz]{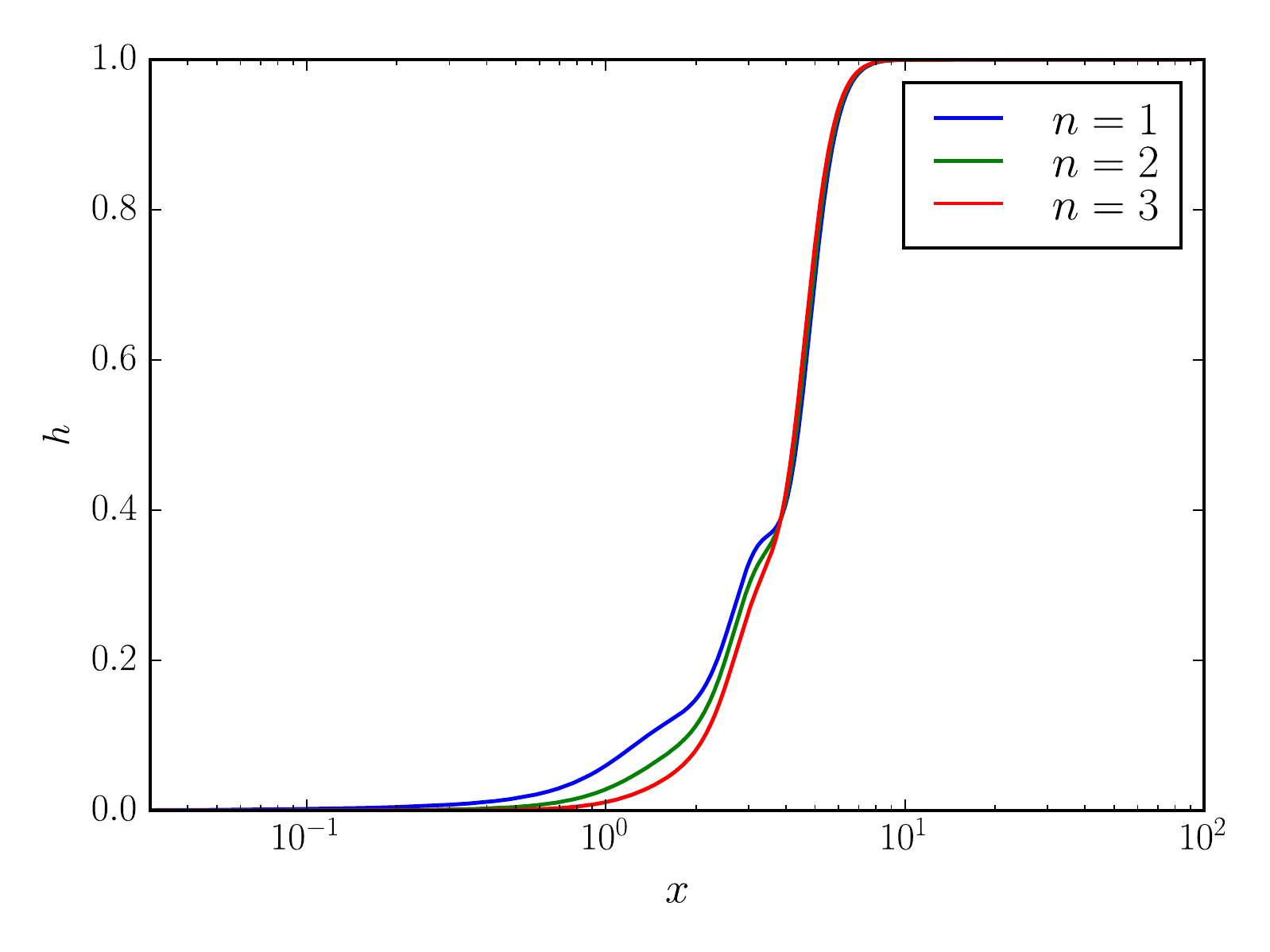}
\includegraphics[width=\fz]{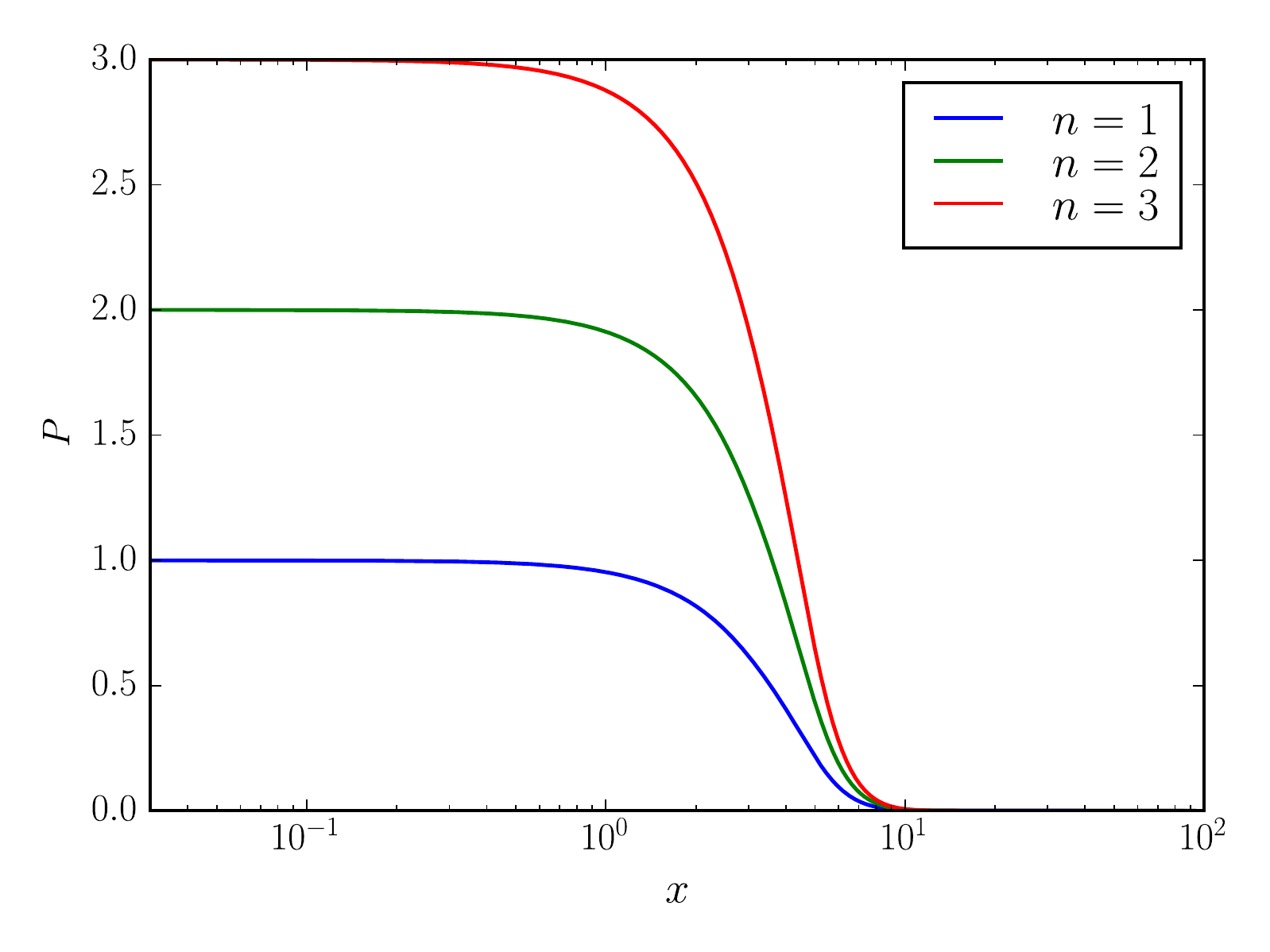}
\caption{The profiles of the condensate, Higgs and gauge field functions
(from top to bottom) for $m=2$, $\alpha_1=1$, $\gamma_2=1$,
$\gamma_3=10$, $q=0.1$, $\tilde{w}=-4$ and winding $n=1, 2, 3$.  }
\label{fig:higher_n}
\end{figure}

Our results for a solution with $m=2$, $\alpha_1=1$, $\gamma_2=1$,
$\gamma_3=10$, $q=0.1$, $\tilde{w}=-4$ are shown in
Fig.~\ref{fig:higher_n}. The condensate function $f(x)$ is practically
unchanged, although we observe a small decrease in the central value
$f(0)$ with increasing $n$ (see Table II for the numerical values).
Moreover, the oscillations in the Higgs field function $h(x)$ that we
observed for $n=1$ persist for $n=2, 3$, although slightly modified. As
far as the integrated quantities are concerned, we observe that the
energy per unit length $\tilde{U}$ {\it per winding $n$}, i.e.
$\tilde{U}/n$ slightly decreases indicating that for our choice of
couplings a superconducting string with higher $n$ can be interpreted as
a {\it bound state} of $n$ superconducting strings with winding $n=1$.
The numerical values of $\tilde{U}/n$ as well as $\tilde{T}/n$ are given
in Table II.  This relates to the observations made in
\cite{Forgacs:2016ndn,Forgacs:2016iva}.  Finally, let us mention that we
also find that the value of the current $\tilde{C}$ decreases with
increasing $n$. A more detailed analysis of this fact is out of the
scope of this paper and is left as future work.

\begin{table}
\begin{tabular}{| c | c | c | c | c | }
  \hline			
  $n$ & $f(0)$ & $\displaystyle\frac{\tilde{T}}{2\pi n}$ &
  $\displaystyle\frac{\tilde{U}}{2\pi n}$ &
  $\displaystyle\frac{\tilde{C}}{2\pi \vert\tilde{v}\vert}$
  \rule{0pt}{2.6ex}\\
  \hline
 $1$ & $1.245$ & $0.314$ & $7.679$ & $1.841$ \\
  $2$ & $1.227$ & $0.234$ & $3.672$ & $1.719$ \\
  $3$ & $1.190$ & $0.234$ & $2.311$ & $1.558$\\
  \hline
\end{tabular}
\label{t2}
\caption{Comparison of some characteristic values of solutions with different windings (see also
 Fig.~\ref{fig:higher_n}). } 
\end{table}

\subsection{Carter stability}

The macroscopic stability criterion of superconducting strings
\cite{Carter:1989xk} relates the velocities of longitudinal and transversal
perturbations to the energy per unit length $U$ and the tension $T$. In
the neutral limit $e_2\to 0$, the definitions above imply that all the
integrated quantities $U$, $T$ and $C$ are positive definite.
One also finds, from the definitions, the useful relationship
\begin{equation}
U-T = |v| C \ \ \ \Longleftrightarrow \ \ \ \ 
\tilde U-\tilde T = |\tilde v| \tilde{C},
\label{UTC}
\end{equation}
from which one can prove \cite{Peter:1992dw} that there exists a finite
neighborhood around $v=0$ for which the string is macroscopically
stable, i.e. both the transverse ($c_\mathrm{_T}$) and the longitudinal
($c_\mathrm{_L}$) velocities, defined above, are real. Indeed, let us
first consider the spacelike case for which $v\geq 0$. In that case,
the energy per unit length happens to equal the Lagrangian from which
one deduces the field equations \eqref{gauge} to \eqref{condensate}, so
that differentiating $U$ with respect to $v$ reduces merely to
differentiating the explicit appearance of $v$. Looking at
Eq.~\eqref{UT}, one sees that this amounts to
\begin{equation}
v\geq 0: \ \ \ \ \ \frac{\dd U}{\dd v} = C \ \ \ \
\underset{\eqref{UTC}}{\Longrightarrow} \ \ \ \ \frac{\dd T}{\dd v} =
-v \frac{\dd C}{\dd v}.
\label{dUdnu}
\end{equation}
Similarly, for $v\leq 0$, the Lagrangian yielding the field equations
now being $T$, one obtains,
\begin{equation}
v\leq 0: \ \ \ \ \ \frac{\dd T}{\dd v} = C \ \ \ \
\underset{\eqref{UTC}}{\Longrightarrow} \ \ \ \ \frac{\dd U}{\dd v} =
-v \frac{\dd C}{\dd v}.
\label{dTdnu}
\end{equation}
We noted earlier that $C\geq 0$, and given its definition
\eqref{C}, it is clear that $\lim_{v\to 0} C=0$: this implies
that for $v\geq 0$, there exists a finite neighborhood around $v=0$ such
that $\dd C/\dd v \geq 0$. In this region, the first equality
in Eq.~\eqref{dUdnu} ensures that $\dd T/\dd v \leq 0$, which, combined
with the second one stating that $\dd U/\dd v \geq 0$, implies that
$c_\mathrm{_L}^2 \geq 0$. Reverting a few signs and using \eqref{dTdnu}
for $v\leq 0$ shows the same conclusion holds in a finite neighborhood
for negative $v$. These arguments depending only on the definition of
the integrated quantities and on the equations of motion that are
satisfied by the fields together with the boundary conditions, show that
there must exist a finite region of state parameter in which the ground
state and the excited configurations are Carter stable for both electric
(timelike) and magnetic (spacelike) currents.

In the region of parameter space studied in Sec.~\ref{StudyNum} however,
the condensate exists only for strictly negative values of the state
parameter, and therefore the argument cannot apply, although it does
apply in many other regimes, such as that discussed in
Ref.~\cite{Hartmann:2016axn}. Here, one must resort to the numerical
solution, such as that shown in Fig. \ref{fig:UTC}. 
We see that Carter criterion for stability is indeed
fulfilled, so it would appear our modes are macroscopically stable. We
must therefore now move on to a local analysis to show the microscopic
instability leading to the cosmological consequences drawn in
Ref.~\cite{Hartmann:2016axn} and further elaborated in our concluding
section \ref{Conclusions}.

\vspace*{2ex} 

\subsection{Linear stability analysis and decay rate}
\label{sub:stability}

To determine the possible physical effects of the excited solutions, a
crucial piece of information is whether they are stable -- and, if not, what is
the typical time scale of their decay. While a full stability analysis
is beyond the scope of the present paper, useful information can be
obtained from the study of linear perturbations, on which we now
concentrate. As we wish to determine the evolution in time of the
solutions after small perturbations, we need the field equations for
$t$- and $z$-dependent fields. For simplicity, we consider only those
solutions where $\sigma$ and $\A_\mu$ are independent on $\theta$. The
field equations are
\begin{widetext}
\begin{equation}
\pd_t^2 \sigma - \pd_z^2 \sigma - \pd_r^2 \sigma - \frac{1}{r} \s \pd_r
\sigma + 2 \s \pd_{\sigma^*} V = 0,
\end{equation} 
\begin{equation}
\lp \pd_t - \ii \s e \s \A_t \rp^2 \phi - \lp \pd_z - \ii \s e \s \A_z
\rp^2 \phi - \lp \pd_r - \ii \s e \s \A_r \rp^2 \phi - \frac{1}{r^2} \s 
\lp \pd_\theta - \ii \s e \s \A_\theta \rp^2 \phi - \frac{1}{r} \s \lp
\pd_r - \ii \s e \s \A_r \rp \phi + 2 \pd_{\phi^*} V = 0,
\end{equation}
\begin{equation}
\pd_\nu \pd^\nu \A_t - \frac{1}{r} \s \pd_r \A_t - \pd_t \lp \pd_\nu
\A^\nu - \frac{1}{r} \s \A_r \rp + \ii \s \frac{e}{2} \s \lc \phi^* \s
\lp \pd_t - \ii \s e \s \A_t \rp \phi - \phi \s \lp \pd_t + \ii \s e \s
\A_0 \rp \phi^* \rc = 0,
\end{equation}
\begin{equation}
\pd_\nu \pd^\nu \A_z - \frac{1}{r} \s \pd_r \A_z - \pd_z \lp \pd_\nu
\A^\nu - \frac{1}{r} \s \A_r \rp + \ii \s \frac{e}{2} \s \lc \phi^* \s
\lp \pd_z - \ii \s e \s \A_z \rp \phi - \phi \s \lp \pd_z + \ii \s e \s
\A_z \rp \phi^* \rc = 0,
\end{equation}
and 
\begin{equation}
\pd_\nu \pd^\nu \A_r + \frac{2}{r^3} \s \pd_\theta \A_\theta - \pd_r
\pd_\nu \A^\nu + \ii \s \frac{e}{2} \s \lc \phi^* \s \lp \pd_r - \ii \s
e \s \A_r \rp \phi - \phi \s \lp \pd_r + \ii \s e \s \A_r \rp \phi^* \rc
= 0 .
\end{equation}
In the following, in order to keep the equations as simple as possible,
we assume $\phi^* \s \pd_\mu \phi \in \mathbb{R}$ for $\mu \neq \theta$.

Let us assume we have a solution $\phi = \phi^{(0)}$, $\sigma =
\sigma^{(0)}$, $\A_\mu = \A_\mu^{(0)}$ of the form given in
Eqs.~\eqref{eq:anz1} and~\eqref{eq:anz2}. We look for perturbed
solutions of the form  $\phi = \phi^{(0)} + \delta \phi$, $\sigma =
\sigma^{(0)} + \delta \sigma$, $\A_\mu = \A_\mu^{(0)} + \delta \A_\mu$,
where
\begin{align}
\delta \phi(t, r, \theta, z) &= p(r) \exp\left[ \ii \s \lp n \s \theta + \sqrt{\la_1}
\s \eta_1 \s \nu \s t - \sqrt{\la_1} \s \eta_1 \s  \kappa \s z\rp\right] \s, \nonumber \\
 \delta \sigma(t, r, \theta, z) &= s(r) \exp \left\{\ii \s \lc (\omega+\sqrt{\la_1} \s
 \eta_1 \s  \nu) \s t - (k+\sqrt{\la_1} \s \eta_1 \s  \kappa) \s z \rc \right\}
 \s , \nonumber \\
\delta B_\theta (t, r, \theta, z) &= a(r) \exp\left[ \ii \s \lp \sqrt{\la_1} \s
\eta_1 \s  \nu \s t - \sqrt{\la_1} \s \eta_1 \s  \kappa \s z \rp \right] \s, 
\end{align}
$(\nu, \kappa) \in \lp \ii \s \mathbb{R} \rp^2$, and $a$, $p$, $s$ are
three real-valued functions. We work in the gauge $\pd_\mu \delta \A^\mu
= 0$ and assume $\A_r = \A_t = \A_z = 0$. One can easily show that the
resulting system of equations is self-consistent provided the algebraic
relation $\omega \s \nu = k \s \kappa$ is satisfied. When allowing $\nu$
and/or $\kappa$ to be more general complex numbers, $\A_r$, $\A_t$ and
$\A_z$ are sourced by the imaginary part of $\phi^* \s \pd_\mu \phi$ and
can thus not be set to zero, which is why we restrict attention to
perturbation satisfying $\Im\mathrm{m}\left( \phi^* \s \pd_\mu \phi
\right) =0$.

The system to be solved is then
\begin{equation} \label{eq:systpert}
\left\lbrace
\begin{aligned}
& \frac{1}{x} \s \pd_x \lp x \s \pd_x s \rp =
	\left( \tilde{w} + \kappa^2 - \nu^2 + 3 \s \gamma_2 \s f^2 - 
	\gamma_2 \s q^2 + \gamma_3 \s h^2 \right) \s s
	+ 2 \s \gamma_3 \s f \s h \s p, \\
& \frac{1}{x} \s \pd_x \lp x \s \pd_x p \rp =
	\lp - \nu^2 + \kappa^2 + \frac{P^2}{r^2} + 3 \s h^2 - 1 + \gamma_3 \s f^2 \rp p
	+ 2 \s \gamma_3 \s h \s f \s s
	- \frac{h \s P}{x^2} \s a, \\
& x \s \pd_x \lp \frac{1}{x} \s \pd_x a\rp =
	\lp - \nu^2 + \kappa^2 + \alpha^2 \s h^2 \rp a
	- \alpha^2 \s P \s h \s \lp p + p_v \rp,
\end{aligned}
\right.
\end{equation}
\end{widetext}
with the boundary conditions $p(0) = s'(0) = a(0) = 0$ and $p(\infty) =
s(\infty) = a(\infty) = 0$. If there exists $\nu \in \ii \s
\mathbb{R}_-$ such that this system has a solution, then the background
solution is linearly unstable in the sense that it supports
perturbations growing exponentially in time. Finding numerical solutions
to this system is challenging, as its exponentially-growing solutions
make it difficult to reach a satisfactory numerical precision for the
bounded ones we are interested in. However, as explained in
Appendix~\ref{app:insta}, one can already obtain information about the
linear stability of the solution by viewing the Higgs and gauge fields
as nondynamical in the linear analysis, i.e., setting $p = a = 0$. The
system~\eqref{eq:systpert} then reduces to
\begin{equation}
\label{eq:lin_s}
\frac{1}{x} \s \pd_x \lp x \s \pd_x s \rp = \left( \tilde w + \kappa^2 - \nu^2
+ 3 \s \gamma_2 \s f^2 - \gamma_2 \s q^2 + \gamma_3 \s h^2 \right) \s s.
\end{equation}
In the present work, since our main aim is to study the nonlinear
solutions rather than linear perturbations we shall work only with
Eq.~\eqref{eq:lin_s}. A more general stability analysis may be
interesting, but is outside of the scope of the present study; besides,
as we also argue below, since the system exhibits instabilities already
for this limited range of perturbation shapes, it can only be shown to
be even more unstable than what we obtain here.

\begin{figure}
\centering
\includegraphics[width = \fz]{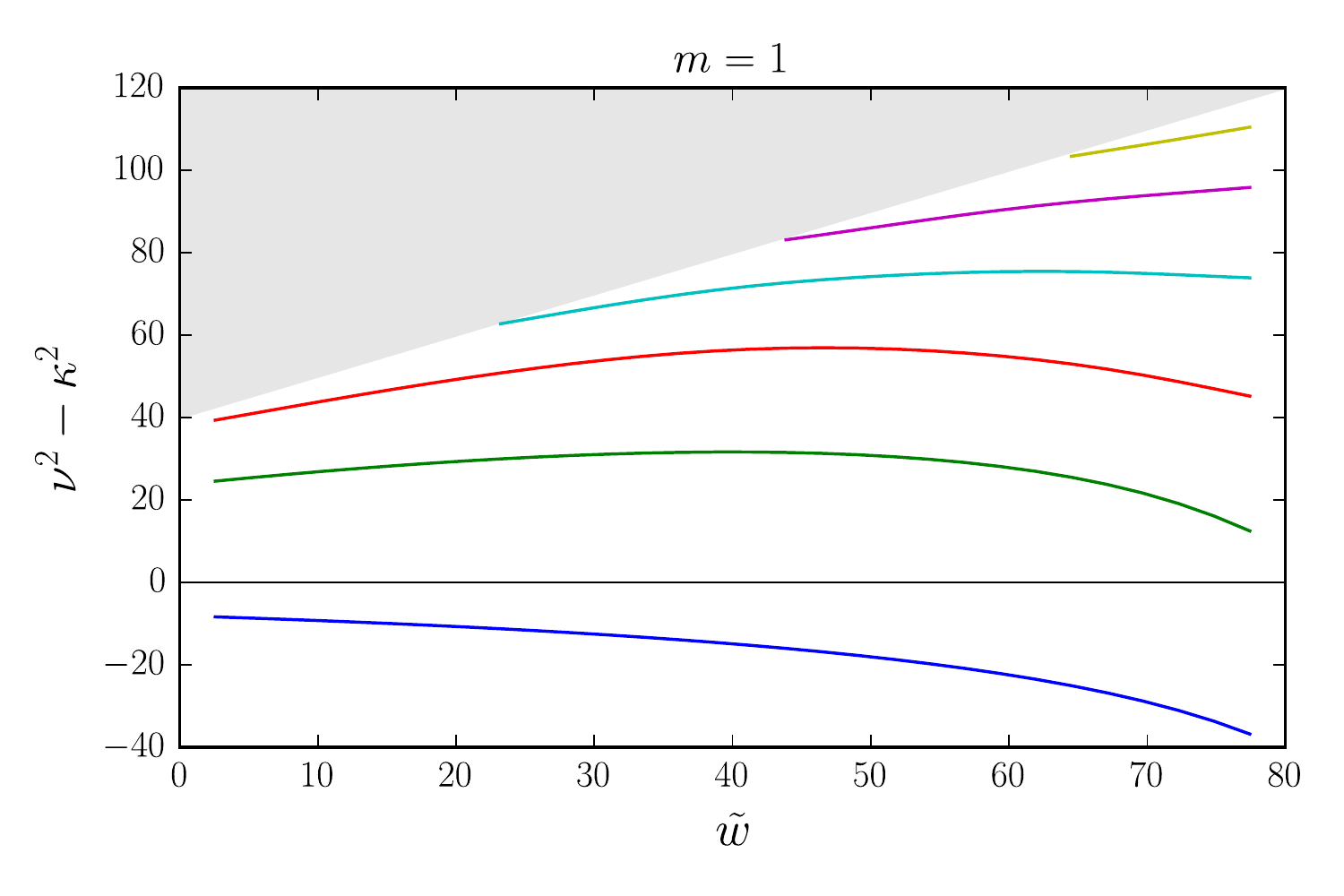}
\includegraphics[width = \fz]{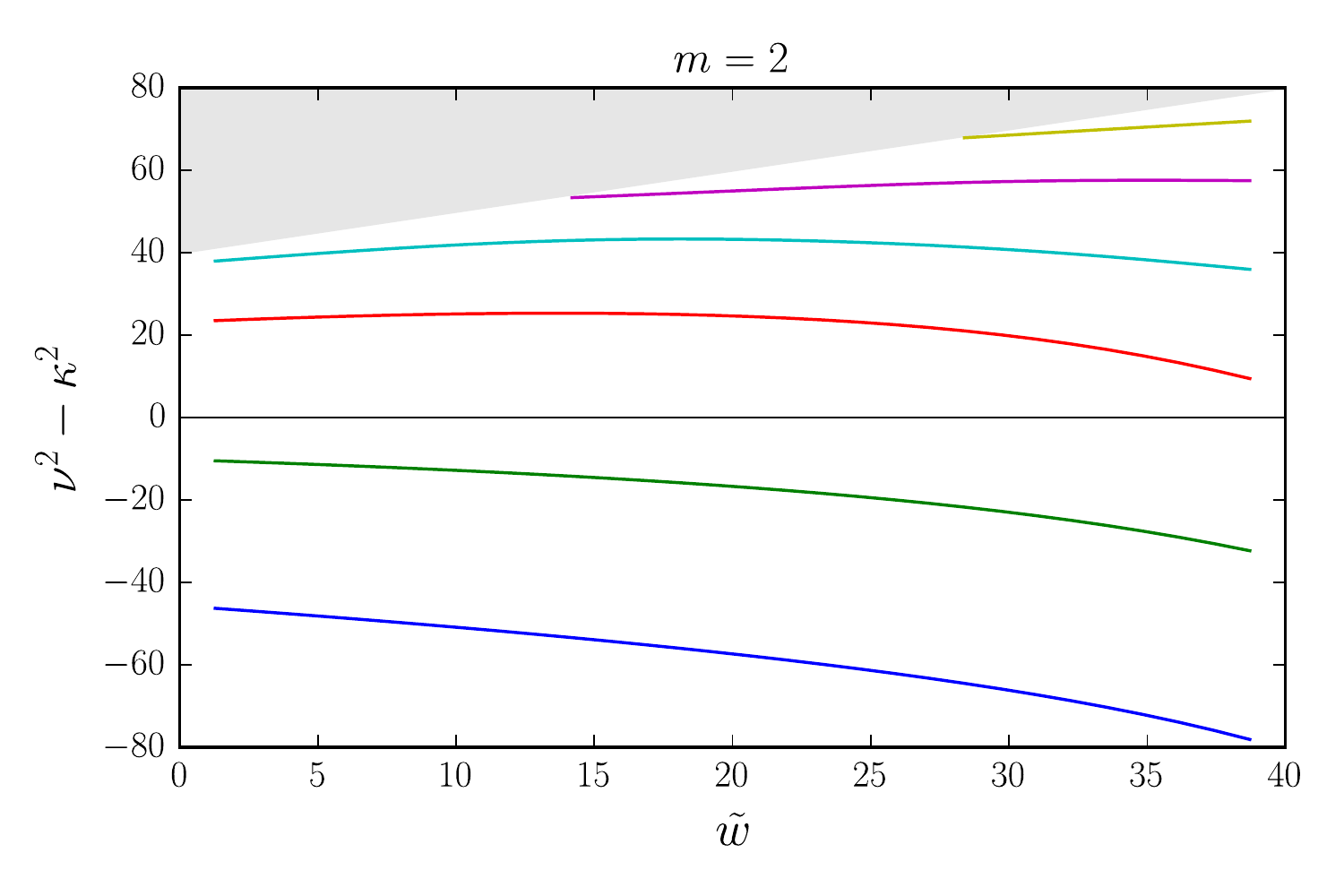}
\includegraphics[width = \fz]{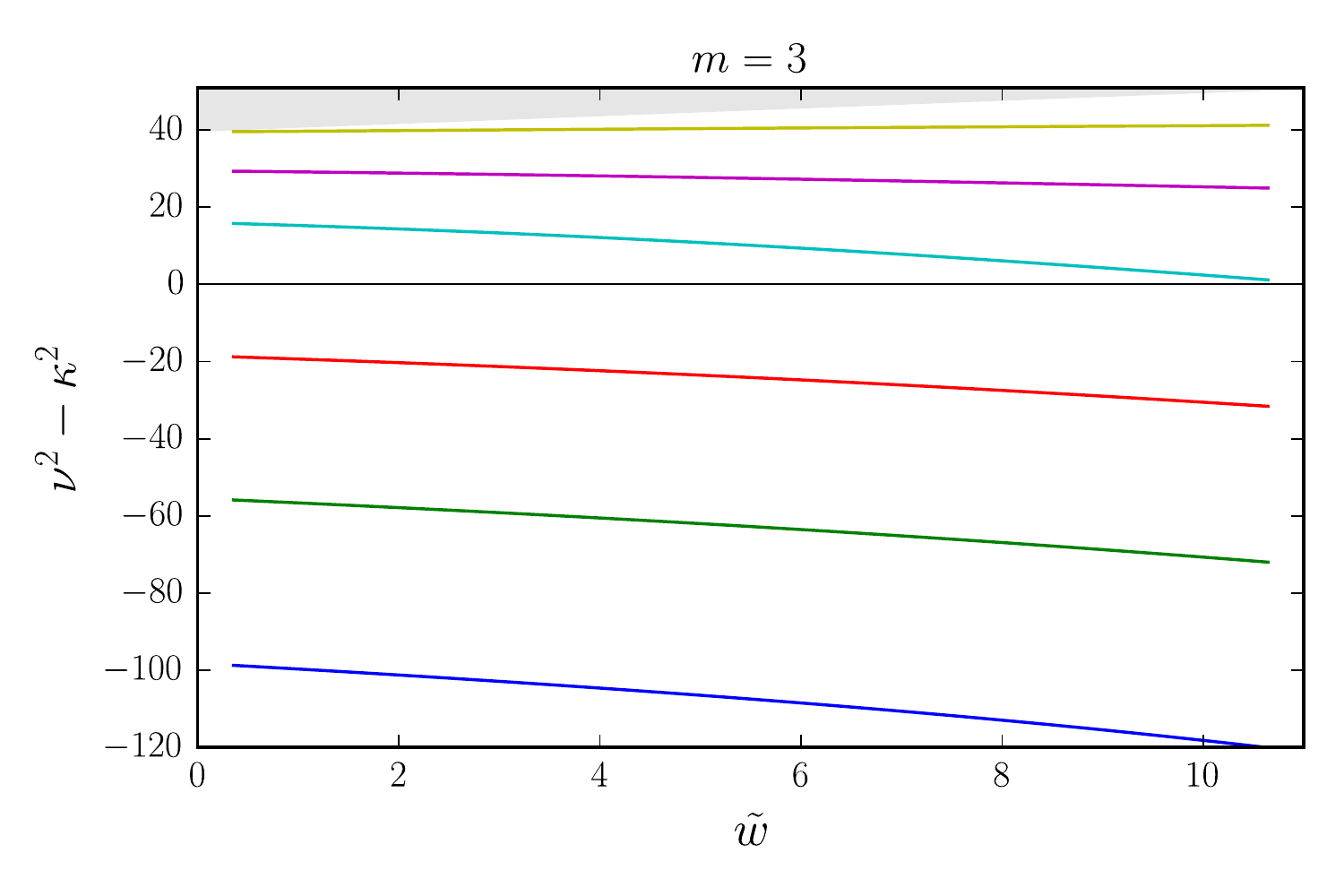}
\caption{Eigenvalues of Eq.~\eqref{eq:lin_s} for the solutions with
$m=1$ (top), $m=2$ (middle), and $m=3$ (top) nodes in the magnetic case
$w > 0$. The parameters are $\gamma_2 = 10$, $\gamma_3 = 200$, and $q =
4$. The background condensate is computed for a Higgs field $h(x) =
\tanh(x)$. The shaded area shows the region $\nu^2 - \kappa^2 > w +
\gamma_3 - \gamma_2 \s q^2$, in which the modes oscillate at infinity
instead of decreasing exponentially. As explained in the text, only
negative values of $\nu^2 - \kappa^2$ correspond to instabilities.}
\label{instab:magnetic}
\end{figure}
\begin{figure*}
\includegraphics[width = \fzz]{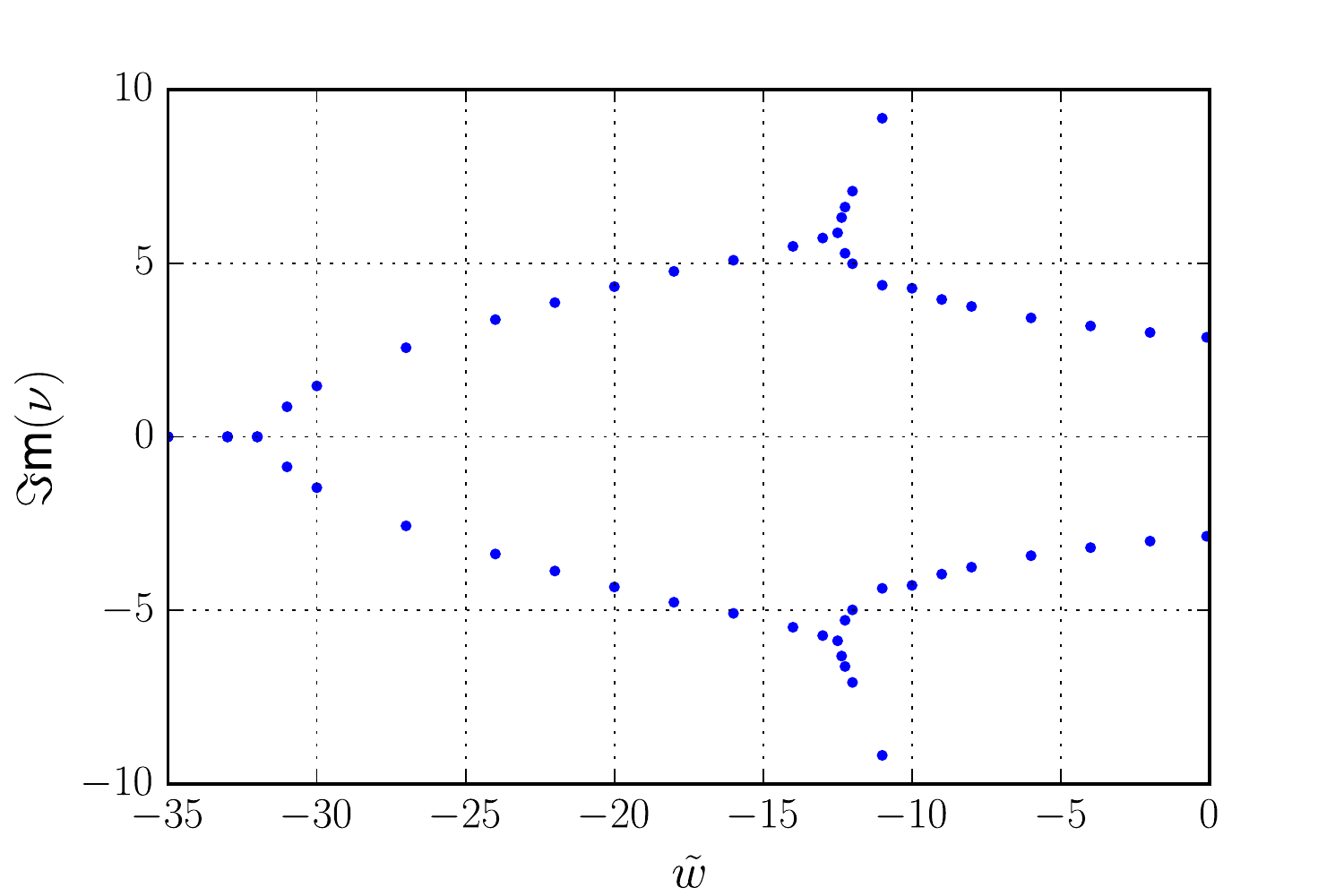}
\includegraphics[width = \fzz]{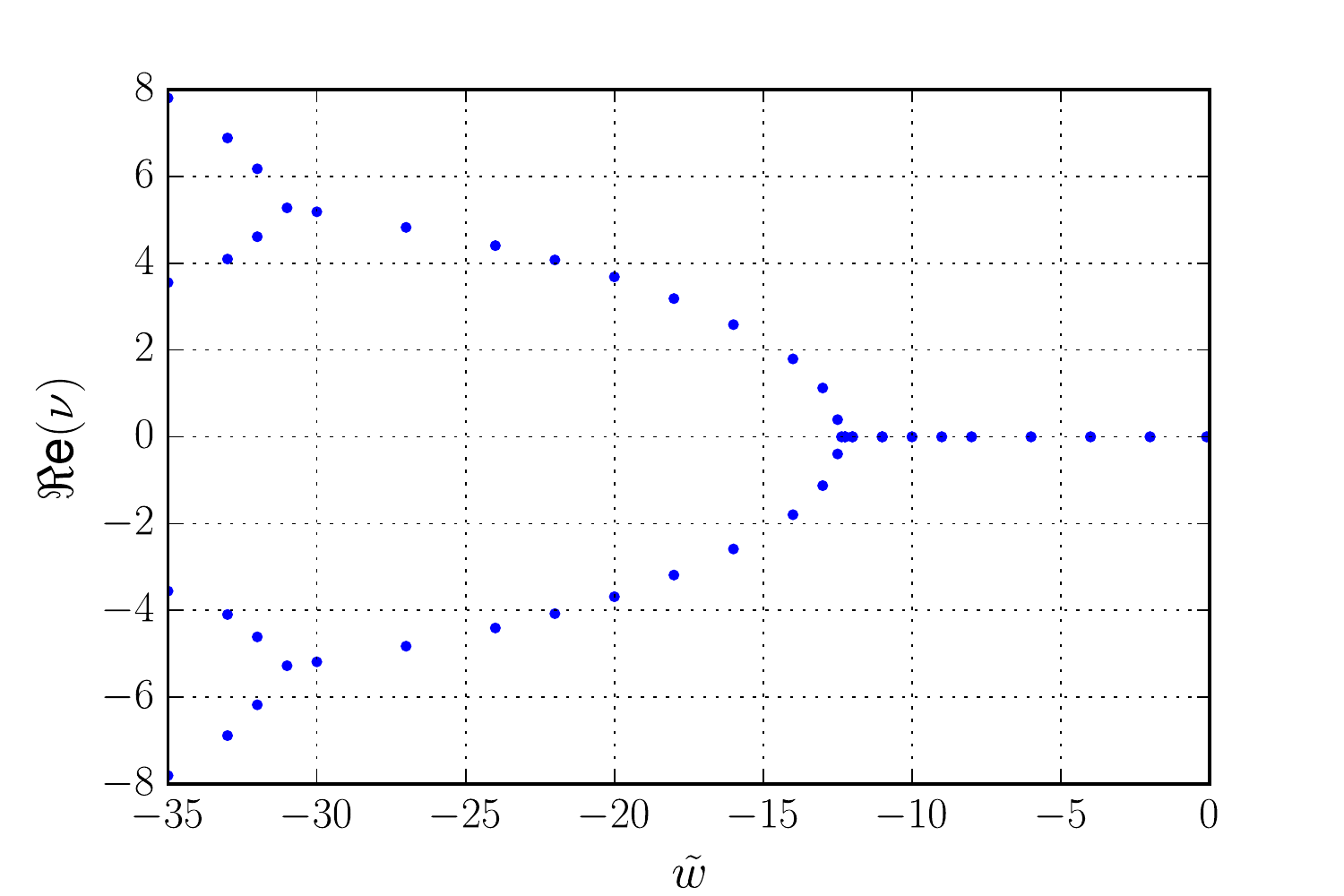}
\caption{Imaginary (left) and real (right) parts of some modes of the
solution with one node in the electric case $w < 0$. The parameters are
$\gamma_2 = 10$, $\gamma_3 = 200$, and $q = 4$. The background
condensate is computed for a Higgs field $h(x) = \tanh(x)$.}
\label{instab:electric}
\end{figure*}

An instability corresponds to a spatially bounded mode growing
exponentially in time (in a given reference frame), i.e., to a bounded
solution of Eq.~\eqref{eq:lin_s} with $\nu^2 - \kappa^2 < 0$. Since the
above derivation requires $\om \s \nu = k \s \kappa$, such solutions
make sense only for the magnetic case $w > 0$. We shall motivate below
that the unstable character of the solutions persists in the case $w <
0$. Fig.~\ref{instab:magnetic} shows the eigenvalues $\nu^2 - \kappa^2$
of Eq.~\eqref{eq:lin_s} for $\gamma_2 = 10$, $\gamma_3 = 200$, and $q =
4$, for the condensates with one, two, and three nodes computed in a
fixed Higgs field background $h(x) = \tanh(x)$. Although only the
solutions with $\nu^2 - \kappa^2 < 0$ yield instabilities, we also show
those with positive values of this quantity to better illustrate what
happens when adding a node to the condensate. The main lessons are the
following:

\begin{itemize}

\item For $\nu^2 - \kappa^2 > \tilde w + \gamma_3 - \gamma_2 \s q^2$,
the solutions oscillate in the large $x$ region, with an amplitude
decaying as $x^{-1/2}$. Bounded solutions thus always exist, providing
the continuous spectrum of Eq.~\eqref{eq:lin_s}.

\item For $\nu^2 - \kappa^2 < \tilde w + \gamma_3 - \gamma_2 \s q^2$, the
solutions are exponentially increasing or decreasing at infinity. When
imposing the boundary condition $s'(0) = 0$, they are thus spatially
bounded only for a discrete set of values of $\nu^2 - \kappa^2$, and
represent the discrete spectrum of Eq.~\eqref{eq:lin_s}.

\item Among these discrete eigenvalues, one, two, and three are negative
for the solutions with one, two, and three nodes, respectively.

\end{itemize}

The third point is the most important one: it means that the solution
with $m$ nodes (for these parameters, and $m$ ranging from $1$ to $3$)
has $m$ unstable modes. This property happens to be satisfied for all
the sets of parameters we tried numerically. We also verified it holds
when working with the actual profile of the Higgs field [solving
Eqs.~\eqref{gauge} -- \eqref{condensate}] instead of the hyperbolic
tangent ansatz. We found no instability for the solutions with $m = 0$.

As mentioned above, the electric case $w < 0$ is more difficult as the
above simplification does not apply~\footnote{The reason is that terms
in $\A_t \s \pd_t \varphi$ and $\A_z \s \pd_z \varphi$ will then appear
in the perturbed Lagrangian, which can thus not be written in the
form~\eqref{eq:formQ}.}. However, since the solutions we found are smooth
in the limit $w \to 0^+$, we conjecture that the aforementioned
instabilities will still be present, at least for small values of $-w$.
To further motivate this, we show in Fig~\ref{instab:electric}
eigenvalues obtained for the condensate with one node, for the same
parameters as in Fig.~\ref{instab:magnetic}. To obtain them, we can no
longer make the assumption $\om \s \nu = k \s \kappa$ and
Eq.~\eqref{eq:lin_s} becomes

\begin{equation}
\begin{aligned}
\lc \tilde w + \gamma_2 \s \lp 2 \s f^2 - q^2 \rp + \gamma_3 \s h^2 - \nu^2 +
\kappa^2 - 2 \s \lp \tilde{\om} \s \nu - \tilde{k} \s \kappa \rp
 \vphantom{\frac{1}{x}} \right. \\ \left.
 - \frac{1}{x} \cdot \pd_x \cdot x \cdot \pd_x \rc s
+ \gamma_2 \s f^2 \s s^* = 0.
\label{eq:s_complex}
\end{aligned}
\end{equation}
We work in a frame where $\tilde{k} = 0$ and look for solutions with
$\kappa = 0$. Notice that the spectrum is invariant under complex
conjugation because Eq.~\eqref{eq:s_complex} is unchanged under $s \to
s^*$, $\nu \to \nu^*$. It is also invariant as well as under the
symmetry transformation $(\tilde{\om}, \nu) \to (-\tilde{\om}, -\nu)$.
As shown in Fig.~\ref{instab:electric}, at least one eigenvalue with a
negative imaginary part is present in most of the domain of $w$ for
which the solution with one node exists. Although the argument of
Appendix~\ref{app:insta} does not apply to this case, this suggests that
these solutions are also unstable. This completes the argument that
excited current-carrying cosmic strings are unstable.

Although we are mostly interested in the case where the winding number
$n$ is equal to $1$, one may wonder if and how choosing a larger value
would affect these results. At the level of the Higgs field, the main
difference lies in the behaviour close to the string axis where $h(x)$
is proportional to $x^n$. To get a first idea of the structure of the
set of solutions for $n > 1$, we thus solved Eq.~\eqref{eq:f2}
numerically in a background field given by
\begin{equation} 
h(x) = \tanh(\kappa \s x)^n, 
\end{equation} 
for $n$ from $2$ and $3$, for the same parameters as in
Fig.~\ref{instab:electric} and with $w = 1$. We obtained similar
results: first, one solution with $m$ nodes exists for $m$ between $0$
and a maximum value (equal to $4$ for $n = 2$ and $5$ for $n = 3$);
second, the solution with $m$ nodes has $m$ unstable modes. We thus
conjecture that the results obtained in this work, concerning both the
structure of excited solutions and stability, remain qualitatively valid
for $n > 1$, as is also confirmed by our numerical construction, shown
in \ref{sub:largern}. A systematic analysis of this case is left for a
future work.

\section{Electromagnetic and gravitational effects}
\label{sec:gauge_and_gravity}

\subsection{Solutions in the U(1)$_{\rm gauge}\times$ U(1)$_{\rm gauge}$
model} \label{sec:gauged}

\begin{figure}[t]
\includegraphics[width = \fz]{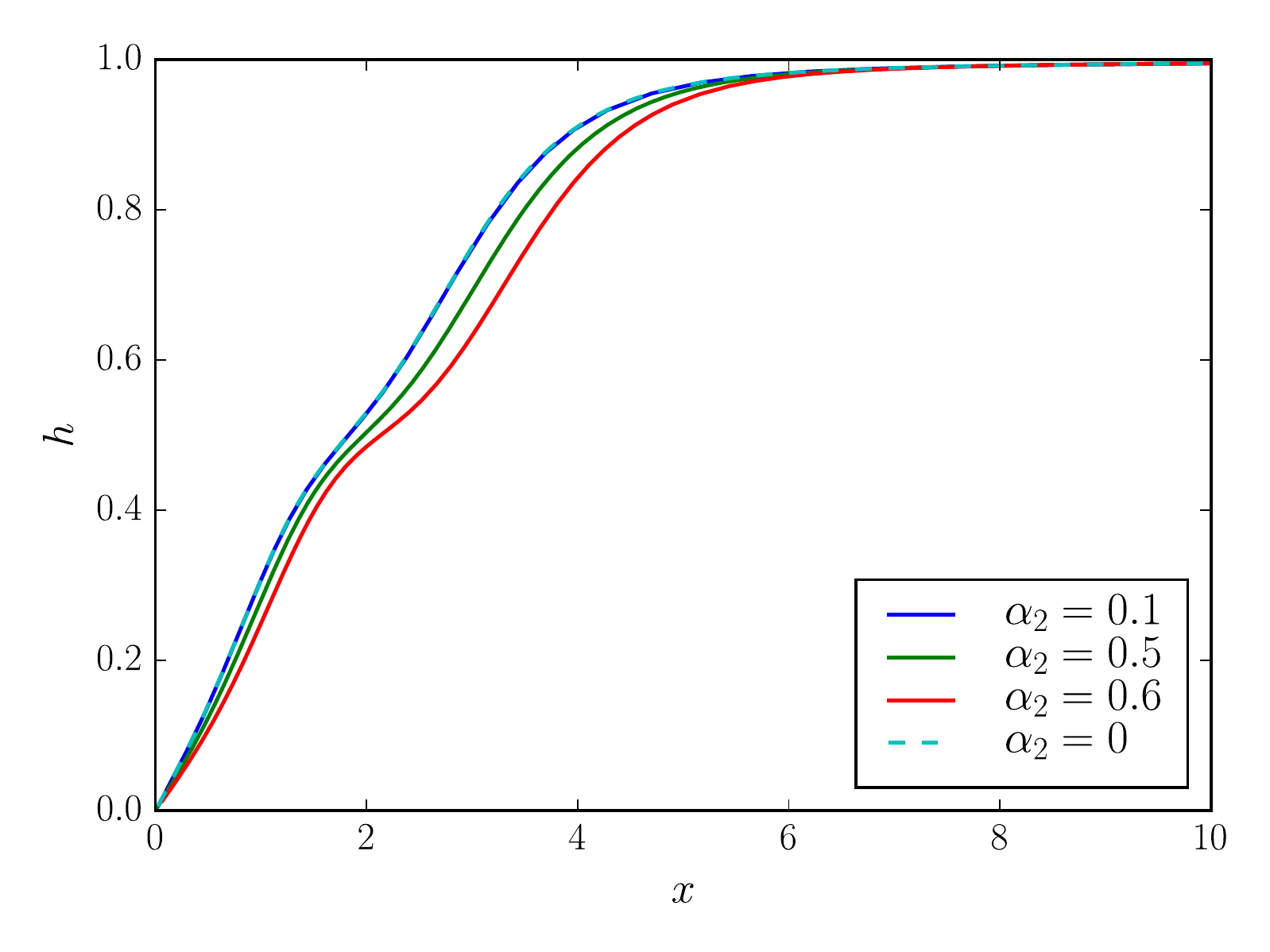}
\includegraphics[width = \fz]{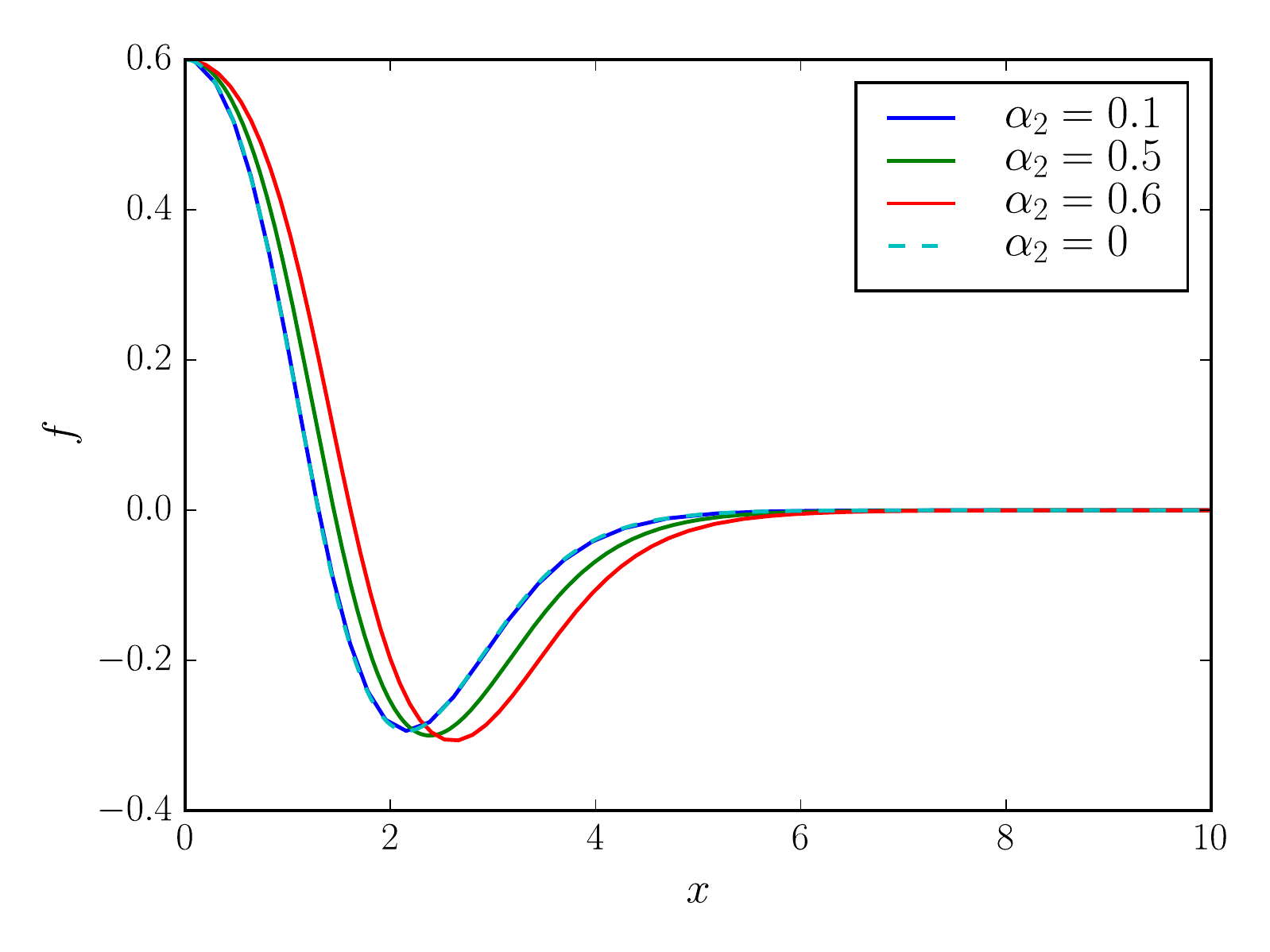}
\includegraphics[width = \fz]{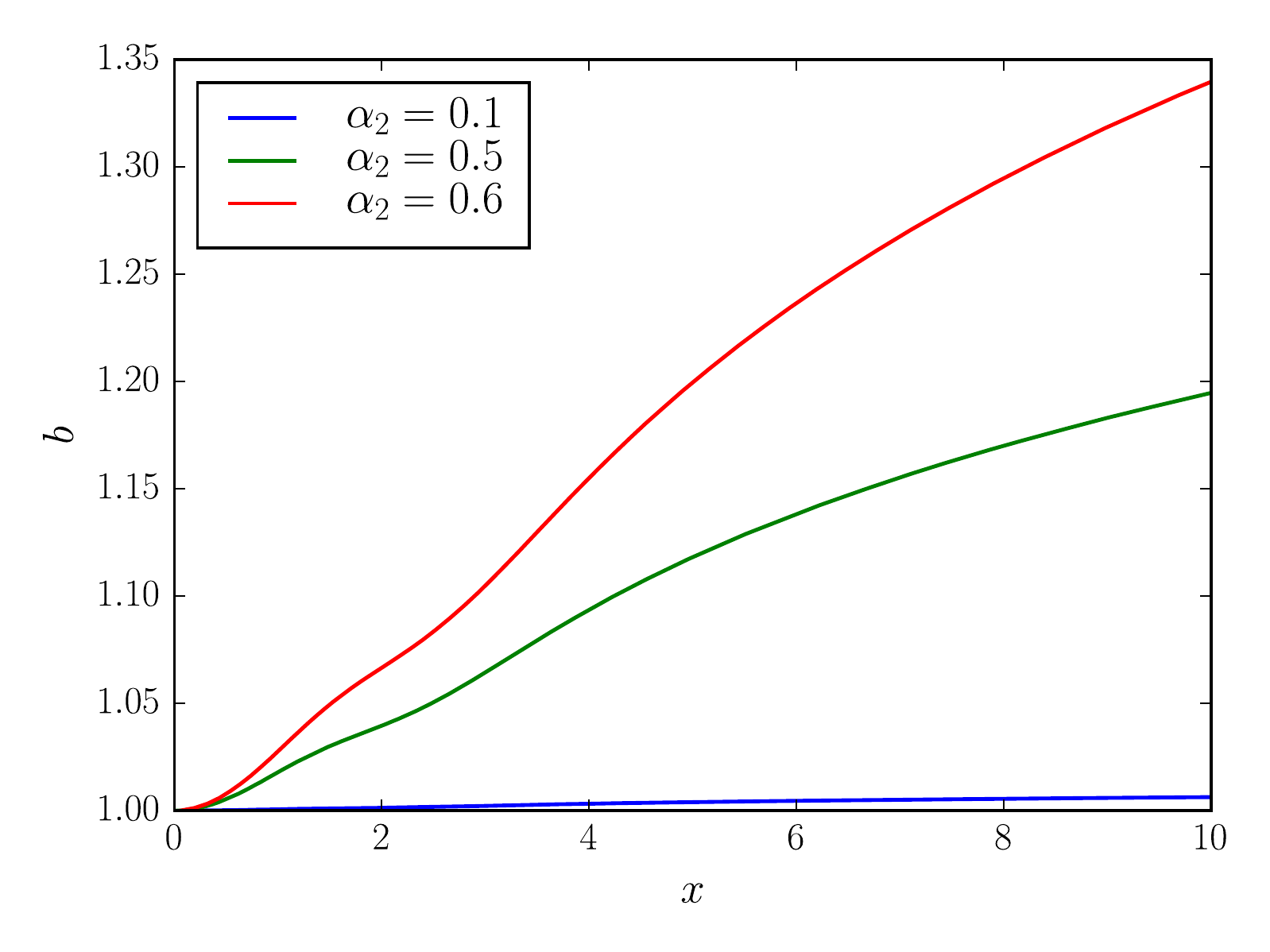}
\caption{Profiles for the string-forming Higgs field $h(x)$ (top panel),
the gauge condensate $f(x)$ (middle) and the asymptotically logarithmic
behaving gauge potential $b(r)-1$ (bottom) for an oscillating radially
excited solution. The different curves correspond to different values of
the rescaled electromagnetic-like coupling constant $\alpha_2$.}
\label{enon0}
\end{figure}

In this subsection, we discuss the effects of the coupling of the
current to an electromagnetic field. Figure \ref{enon0} shows the field
profiles for various values of $\alpha_2$ in the case $m=1$. Similar
results were obtained for various values of $m$, showing that, as for
the background mode \cite{Peter:1992ta}, the internal structure of the
current-carrying cosmic string is essentially not modified by inclusion
of electromagnetic effects, the latter being, if anything, only capable
of long range interactions on the macroscopic behavior of the strings
\cite{Peter:1993mv,Gangui:1997bi}. The figure also shows clearly the
expected behavior of the gauge potential sourced by an infinitely long
current-carrying string, i.e. $b(r)\sim \ln \lp r/r_\sigma\rp$, where
$r_\sigma \simeq m_\sigma^{-1}$, the Compton wavelength of the current
carrier $\sigma$, provides an order of magnitude estimate of the
electromagnetic radius of the vortex.

\subsection{Gravitational effects}
\label{sec:GE}

The space-time of a superconducting string possesses a deficit angle
$\Delta \sim U+T$, similar to that of a Nambu-Goto string
\cite{Linet:1989xj}, while locally there exists an attractive force
towards the string \cite{Moss:1987ka,Peter:1993wx,Peter:1993ww},
potentially leading to observable effects \cite{Garriga:1994sy}.

The existence of a deficit angle is responsible for a number of physical
effects (for a recent review see \cite{Vachaspati:2015cma}). When the
string moves, it creates wakes that could e.g. be observable in the 21cm
radiation from hydrogen, while the so-called Kaiser-Stebbins-Gott effect
\cite{Kaiser:1984iv} leads to discontinuities in the CMB. Furthermore,
the deficit angle would lead to gravitational lensing that is quite
distinct from that caused by other spatially extended objects. Finally,
so-called kinks and cusps on strings as well as the oscillations of
string loops are believed to emit gravitational waves.

In order to discuss gravitational effects, we couple the model
(\ref{lag}) minimally to gravity and choose the following
parametrization of the metric tensor
\begin{equation}
\dd s^2= N^2(x) \dd t^2 - \dd x^2 - L^2(x) \dd \theta^2 - K^2(x) \dd z^2  \ .
\label{metric}
\end{equation}
This model has already been studied in \cite{Hartmann:2012mh} and we
refer the reader for more details to this paper. Let us just remark here
that there is an extra dimensionless coupling in the model, which
corresponds to the ratio between the symmetry breaking scale $\eta_1$
and the Planck mass $M_\mathrm{Planck}=G_\mathrm{N}^{-1/2}$:
\begin{equation}
\beta=8\pi G_\mathrm{N} \eta_1^2,
\end{equation}
with $G_\mathrm{N}$ the Newton constant.

Given the numerical solutions to the coupled matter and gravity
equations, we can read of the deficit angle of the space-time from the
behaviour of the metric function $L(x)$:
\begin{equation}
 \Delta=2\pi(1-c_1), \ \hbox{where} \ \ \ L(x\rightarrow\infty)
 \rightarrow c_1 x +c_2
\end{equation}
where $c_1$ and $c_2$ are constants that have to be determined numerically. 

Solving the coupled matter and Einstein equations numerically
we determined the deficit angle for the oscillating string solutions, which 
is given by the sum of the energy density $U$ and the tension $T$. 
This is nothing new in comparison to the fundamental string solutions,
however, we now have {\it a dicrete set of values of the deficit angle for
one fixed set of coupling constants}. Hence, measuring the deficit
angle, e.g. by gravitational lensing, does not uniquely determine the values of
the couplings in the model.

\begin{figure}
\includegraphics[width = \fz]{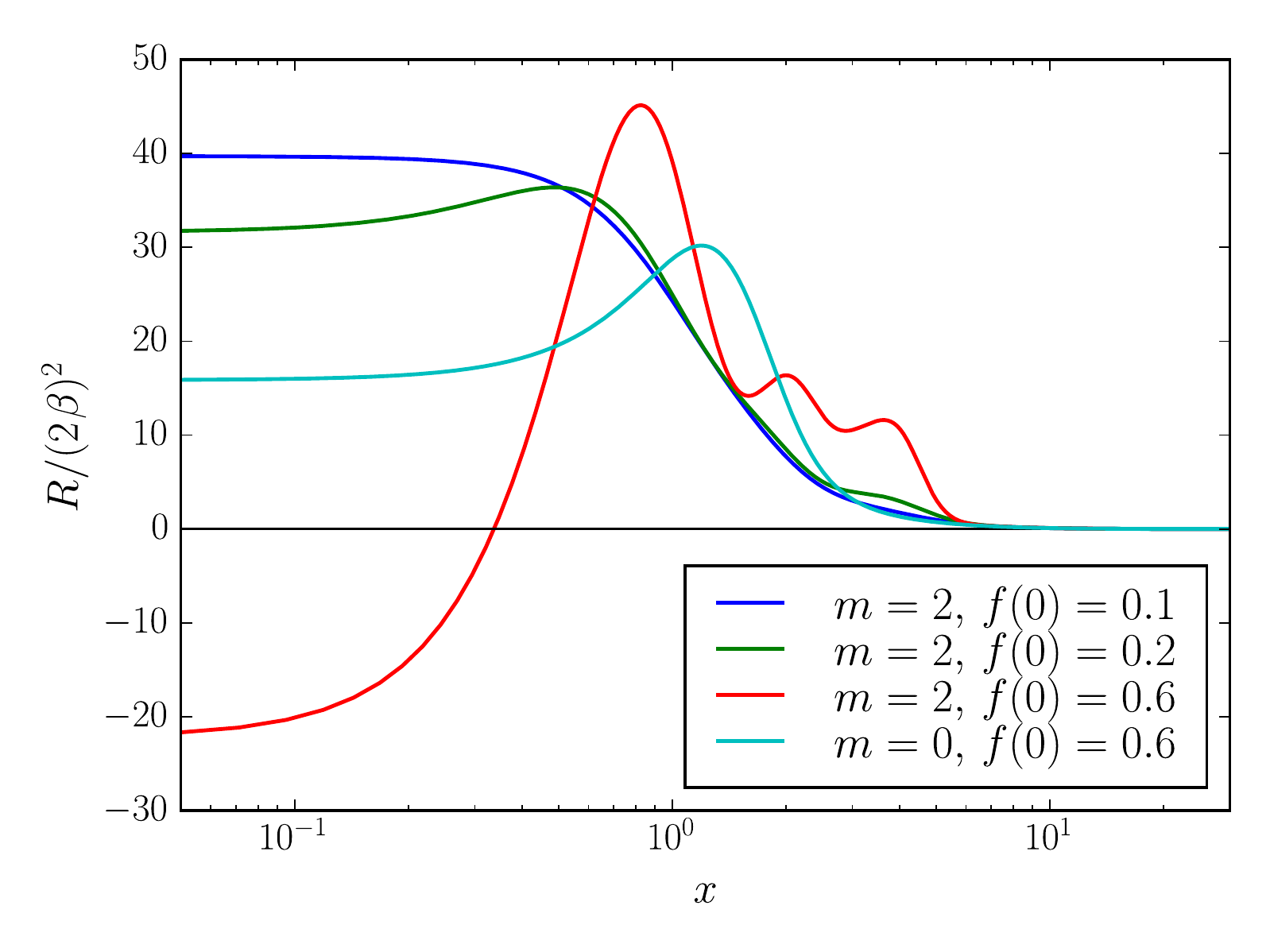}
\caption{Ricci scalar as a function of string core distance for an
$m=2$ oscillatory mode with various values of the condensate
interior value $f(0)$.}
\label{fig:RicciPlot}
\end{figure}

We also observe a new effect that is related to the oscillations of the
Higgs field appearing for sufficiently large values $f(0)$. We find that
these trigger an oscillation in the local scalar curvature. This is
demonstrated in Fig.~\ref{fig:RicciPlot} for a $m=2$ solution and
various values of the condensate.

\section{Conclusions}
\label{Conclusions}

In this paper, we have studied excited cosmic string solutions with
superconducting currents. These solutions possess a number of nodes in
the condensate field function and can trigger -- for sufficiently large
condensates -- oscillations in the Higgs field function as well as in
the local scalar curvature in the space-time around the string. Though
some of these solutions are macroscopically, i.e. Carter stable, we show
that they are microscopically unstable and would decay rapidly after
formation.

In the macroscopic description of cosmic strings, which characterizes
them solely in terms of their energy per unit length $U$ and tension
$T$, our results are interesting because they imply that for a given set
of physical parameters, a discrete number of cosmic strings with
different values of $U$ and $T$ exist. Assuming that at the formation of
cosmic string networks in the primordial universe these excited
solutions can be formed, the evolution of the network would involve
(from a macroscopic point of view) a number of different types of
strings, of which some are unstable. Certainly, this will modifiy the
dynamics and evolution of string networks and the question arises
immediately whether and how these networks reach a scaling solution.

Moreover, the gravitational effects of cosmic strings are determined by
the deficit angle in their space-time (which in turn is determined
solely by $U$ and $T$), leading to a number of observable effects such
as gravitational lensing as well as wakes and the Kaiser-Stebbins-Gott
effect. Now since strings with different values of $U$ and $T$ exist,
these will lead to different effects e.g. in the Cosmic Microwave
background (CMB) spectra.

From a microscopic point of view, the instability of excited solutions
leads to emission of high energy particle radiation that could e.g. be
observed in the form of cosmic rays. Moreover, since the local curvature
of the space-time around the string is modified by the condensate, it is
conceivable that when decaying an additional emission of primordial
gravitational waves can be expected.

Finally, let us state that our analysis clearly shows that the
underlying field theoretical structure plays a very important role --
even when considering only the macrophysics and, hence, integrated
quantities.  For Nambu--Goto simulations of cosmic string network
evolution (see e.g.
\cite{Bennett:1989yp,Bennett:1989ak,BlancoPillado:2011dq}) the existence
of a network of strings with different tensions and the emission of
gravitational waves from excited strings could be of relevance, while
for Abelian--Higgs string simulations (see e.g.
\cite{Vincent:1997cx,Moore:2001px,Hindmarsh:2008dw} as well as
\cite{Hindmarsh:2011qj} and references therein) the excitations of the
Higgs field as well as the existence of high energy particle radiation
could be interesting to take into account.

\acknowledgments

We thank Árpád Lukács for his careful reading of the manuscript and his
suggestions. BH would like to thank FAPESP for financial support under
grant number {\it 2016/12605-2} and CNPq for financial support under
{\it Bolsa de Produtividade Grant 304100/2015-3}.  PP would like to
thank the Labex Institut Lagrange de Paris (reference ANR-10-LABX-63)
part of the Idex SUPER, within which this work has been partly done.

\appendix

\section{A note on instabilities}
\label{app:insta}

In this appendix we show that, under some conditions, it is possible to
study the stability of a field configuration without solving the full
set of linearized field equations. More precisely, assuming the field
theory has a Hamiltonian structure and the boundary conditions are such
that the relevant operator can be diagonalized, we show that finding one
unstable mode when viewing all the fields except one ~\footnote{This
result can be easily extended to an arbitrary number of dynamical fields
following the same steps. For conciseness, we restrict here to the case
of one single dynamical field, used in the main text.} as nondynamical
implies that the full theory, with all fields dynamical, also has an
instability. Moreover, the growth rate of perturbations in the
``restricted'' problem with only one dynamical field gives a lower bound
on the growth rate of the most unstable mode in the ``full'' problem. We
first focus on the simpler case of a classical particle in two
dimensions, which provides some intuition as to why adding one degree of
freedom generally does not make a system more stable. We then generalize
the results to an arbitrary finite number of dimensions and to field
theory. Finally, we explain why they apply to the model dealt with in
the main text.

\subsection{A toy-model: Classical point particle in a 2D potential}

Let us consider a classical particle with mass $m > 0$ in a
two-dimensional space, subject to a potential $V$. To make things
simple, let us assume $V$ is quadratic:
\begin{equation}
V(x,y) = \frac{A}{2} \s x^2 + \frac{B}{2} \s y^2 + C \s x \s y,
\end{equation}
where $A$, $B$, and $C$ are three real numbers. The equation of motion
is
\begin{equation} \label{eq:2Newton}
m \s \pd_t^2 
\begin{pmatrix}
x \\ y
\end{pmatrix}
= 
- \begin{pmatrix}
A & C \\
C & B
\end{pmatrix}
\begin{pmatrix}
x \\ y
\end{pmatrix}
.
\end{equation}
To determine the stability of the equilibrium position $x = y = 0$, one
can look for solutions whith $(x, y) \propto \e^{\ii \s \nu \s t}$ with
$\nu \in \mathbb{C}$: the equilibrium is stable if all possible values
of $\nu$ are real, and unstable otherwise. Plugging this ansatz into
Eq.~\eqref{eq:2Newton}, one finds nontrivial solutions exist if and only
if
\begin{equation}
\abs{\begin{matrix}
A - m \s \nu^2 & C \\
C & B - m \s \nu^2
\end{matrix}}
= 0. 
\end{equation}
The eigenvalue equation is thus:
\begin{equation}
(A - m \s \nu^2) \s (B - m \s \nu^2) - C^2 = 0. 
\end{equation}
A straightforward calculation gives the possible eigenvalues as
\begin{equation}\label{eq:eigen2}
\nu^2 = \frac{(A+B) \pm \sqrt{(A-B)^2 + 4 \s C^2}}{2 \s m} . 
\end{equation}
Although it is easy from this expression to determine directly the
stability condition, we will here follow a different route which will be
easier to generalize to a larger number of degrees of freedom.

This two-dimensional case is very particular in that all eigenvalues can
be computed explicitly. However, this is generally not the case in the
presence of a large number of degrees of freedom. One possible way to
simplify the calculations is to assume some of them are not dynamical.
In the present case, for instance, one could set by hand $y = 0$ and
consider only the stability in the $x$ direction. Then, the equilibrium
position will be stable if $A > 0$ and unstable if $A < 0$. Moreover, in
the latter case the growth rate is: $\mathrm{Im}(\nu) = \sqrt{-A / m}$.

Let us now return to Eq.~\eqref{eq:eigen2} and see what the condition $A
< 0$ for instability of the ``restricted'' problem with $y$ set to $0$
by hand can tell us about the stability of the ``full'' problem where
$x$ and $y$ are both dynamical. Taking the $-$ sign in this equation,
one gets
\begin{equation}
\begin{aligned}
\nu^2 & = \frac{(A + B)- \sqrt{(A-B)^2 + 4 \s C^2}}{2 \s m} \leq
\frac{(A + B)- (A-B)}{2 \s m} \\ \nu^2 & \leq \frac{A}{m}.
\end{aligned}
\end{equation}
So, the ``full'' problem also shows an instability, with a growth rate
larger than or equal to $\sqrt{-A/m}$. This illustrates a general fact:
instabilities obtained when freezing some degrees of freedom give a
lower bound on the growth rate of the strongest instability in the
``full'' problem.

\subsection{Generalization to a finite number of degrees of freedom}

Let us generalize this to any finite number $N$ of real degrees of
freedom. Let $\Phi$ be the vector of perturbations with respect to some
equilibrium point. We assume the Lagrangian has the form
\begin{equation} \label{eq:pertL_finitedof}
L = L^{(0)} + \frac{1}{2} \s (\pd_t \Phi)^T \s (\pd_t \Phi) -
\frac{1}{2} \s \Phi^T \cdot K \cdot \Phi + O(\Phi^3),
\end{equation}
where $L^{(0)}$ is evaluated at the equilibrium point (and thus
independent on $\Phi$), a superscript $T$ denotes vector transposition,
and $K$ is a real matrix. The term $O(\Phi^3)$ denotes higher-order
terms. Without loss of generality, one can assume $K$ is
symmetric, since if $K$ is not symmetric, one can replace it with
its symmetric part $(K + K^T)/2$, which does not change the value of
$L$. Neglecting higher-order terms in $\Phi$, the evolution equation
for perturbations is:
\begin{equation}
\pd_t^2 \Phi = - K \cdot \Phi. 
\end{equation}
Solutions with $\Phi \propto \e^{\ii \s \nu \s t}$ exist if and only if
$\nu^2$ is an eigenvalue of $K$.

The above analysis can be straightforwardly generalized to complex
degrees of freedom by separating their real and imaginary parts,
provided the perturbed Lagrangian can be written in the
form~\eqref{eq:pertL_finitedof} with vector transposition replaced by
hermitian conjugation. The operator $K$ can then be chosen to be
hermitian without loss of generality.
 
Let us define the ``restricted'' problem by assuming that only the $M <
N$ first degrees of freedom are dynamical. We denote wih a superscript
$(R)$ quantities pertaining to the ``restricted'' problem. So,
$\Phi^{(R)}$ denotes the vector of the $M$ first components of $\Phi$
and $K^{(R)}$ the $M$ by $M$ submatrix of $K$ obtained by taking only
the first $M$ lines and columns. The vector $\Phi^{(R)}$ then obeys the
equation:
\begin{equation}
\pd_t^2 \Phi^{(R)} = - K^{(R)} \cdot \Phi^{(R)}. 
\end{equation}
Let us assume the ``restricted'' problem has an instability, i.e., that
$K^{(R)}$ has a strictly negative eigenvalue $\lambda^{(R)}$. Then there
exists a configuration $\Phi^{(R)}_0 \neq 0$ such that
\begin{equation}
K^{(R)} \cdot \Phi^{(R)}_0 = \lambda^{(R)} \s \Phi^{(R)}_0. 
\end{equation}
Our goal is to show that $K$ also has a strictly negative eigenvalue
$\la$, such that $\la \leq \lambda^{(R)}$. This will prove that the
``full'' problem is also unstable, with a growth rate larger than or
equal to that of the ``restricted'' problem.

We proceed by contradiction. Let us assume for a moment that all
eigenvalues $\la_i$ of $K$ are strictly larger than $\la^{(R)}$. Since
$K$ is a symmetric real matrix, it can be diagonalized in an orthonormal
basis. Let $\Phi$ be any non-vanishing vector. Let us expand it as 
\begin{equation}
\Phi = \sum_i a_i \s \Phi_i,
\end{equation}
where $(\Phi_i)_{1 \leq i \leq N}$ is an orthonormal basis of
eigenvectors of $K$, such that $K \s \Phi_i = \la_i \s \Phi_i$, and the
$a_i$ are real numbers. We have:
\begin{align}
\Phi^T \cdot K \cdot \Phi & = \sum_i a_i^2 \s \la_i > \sum_i a_i^2 \s
\la^{(R)} \\ \Phi^T \cdot K \cdot \Phi & > \la^{(R)} \s \Phi^T \cdot
\Phi.
\end{align} 
To get a contradiction, we thus only have to find a vector for which
this inequality is not satisfied. One example of such a vector,
$\Phi_0$, is found by taking the first $M$ components of $\Phi_0^{(R)}$
and $N-M$ zeros. We write it schematically as
\begin{equation}
\Phi_0 \equiv 
\begin{pmatrix}
\Phi_0^{(R)} \\
0
\end{pmatrix}.
\end{equation}
Then,
\begin{equation}
K \cdot \Phi_0 = 
\begin{pmatrix}
K^{(R)} \cdot \Phi_0^{(R)} \\
*
\end{pmatrix} = 
\begin{pmatrix}
\lambda^{(R)} \s \Phi_0^{(R)} \\
*
\end{pmatrix},
\end{equation}
where the star represents $N-M$ coefficients which play no role in the
following, so that
\begin{equation}
\Phi_0^T \cdot K \cdot \Phi_0 = 
\la^{(R)} \s \Phi_0^{(R) T} \cdot \Phi_0^{(R)} =
\la^{(R)} \s \Phi_0^{T} \cdot \Phi_0. 
\end{equation}
We obtain a contradiction, which shows that $K$ has at least one
eigenvalue smaller than or equal to $\la^{(R)}$.

\subsection{Generalization to a field theory}

Let us consider a theory with $N \in \mathbb{N}^*$ real fields $\psi_i$,
$i \in [\![1, N]\!]$, in $(d+1)$ dimensions. For all $i \in [\![1,
N]\!]$, we denote by $\phi_i$ a perturbation of the field $\psi_i$. We
define the vector
\begin{equation}
\Phi \equiv 
\begin{pmatrix}
\phi_1 \\ \phi_2 \\ \vdots \\ \phi_N
\end{pmatrix} .
\end{equation}

Let us assume that the quadratic action $\mathcal{S}^{(2)}$ may be written as
\begin{equation}
\mathcal{S}^{(2)} = \int \dd t \s \dd^d x \s \sqrt{\abs{g}} \s \lp
\frac{1}{2} \s \pd_t \Phi^T \cdot \pd_t \Phi
- \frac{1}{2} \s \Phi^T \cdot K \cdot \Phi
\rp ,
\end{equation}
where $K$ is a real matrix of differential operators which does not
involve $\pd_t$ and is independent of $t$, and where $g$ is the
determinant of the metric, assumed to be be everywhere nonvanishing. We
assume all functions and their derivatives are bounded. As above, a
superscript ``$T$'' indicates vector transposition. As above also,
without loss of generality, one can assume $K$ is symmetric for the
$L^2$ scalar product. Let us further assume that $g$ is independent on
time. The linear equation on perturbations is then
\begin{equation}\label{eq:nret}
\pd_t^2 \Phi = - K \cdot \Phi.
\end{equation}
For each negative eigenvalue $\la$ of $K$, there is thus a growing and a
decaying mode in time, as $\e^{\pm \sqrt{- \la} \s t}$. Conversely, any
mode growing or decaying exponentially in time with rate $\nu$
corresponds to a negative eigenvalue $- \nu^2$ of $K$.

Let us assume that the ``restricted'' equation
\begin{equation}\label{eq:res}
\pd_t^2 \phi_1 = K_{1,1} \s \phi_1,
\end{equation}
where $K_{1,1}$ denotes the $(1,1)$ component of $K$, has a strictly
negative eigenvalue $\la_0$. Then there exists a nonvanishing solution
$\phi_1^{(0)}$ such that
\begin{equation}
K_{1,1} \s \phi_1^{(0)} = \la_0 \s \phi_1^{(0)}.
\end{equation}
Let us define the following vector of functions with only one
nonvanishing component:
\begin{equation}
\Phi^{(0)} \equiv 
\begin{pmatrix}
\phi_1^{(0)} \\ 0 \\ \vdots \\ 0
\end{pmatrix} .
\end{equation}
We have:
\begin{align}
\int  \dd^d x \s \sqrt{\abs{g}} \s \Phi^{(0) T} \cdot K \cdot \Phi^{(0)} = {} &
\int  \dd^d x \s \sqrt{\abs{g}} \s \phi_1^{(0)} \s K_{1,1} \s \phi_1^{(0)} 
\nonumber \\
= {} & \la_0 \s \int  \dd^d x \s \sqrt{\abs{g}} \s \lp \phi_1^{(0)} \rp^2 < 0.
\end{align}
Since $K$ is real and symmetric, it is hermitian and thus
diagonalizable. From the above expression, using the same argument as in
the case of finite number of dimension, one deduces that it has at least
one strictly negative eigenvalue (otherwise bracketing it with a $L^2$
vector could give only positive or vanishing values). Moreover, for any
$\la_1 > \la_0$, the same argument applies to $K - \la_1 \s \mathbf{1}$,
where $\mathbf{1}$ is the identity operator, showing that $K - \la_1 \s
\mathbf{1}$ has (at least) one strictly negative eigenvalue, and thus
that $K$ has one eigenvalue strictly smaller than $\la_1$. Taking the
limit $\la_1 \to \la_0$, one finds that $K$ has (at least) one
eigenvalue smaller than or equal to $\la_0$.

So, under the hypotheses of this subsection, the existence of a strictly
negative eigenvalue $\la_0$ for the ``restricted''
problem~\eqref{eq:res} implies that of (at least) one strictly negative
eigenvalue $\la_0' \leq \la_0$ for the full problem~\eqref{eq:nret}.

This argument can be made manifestly Lorentz-invariant in the (t,z)
plane by considering an action of the form:
\begin{align}\label{eq:formQ}
& \mathcal{S}^{(2)} = \int \dd t \s \dd^d x \s \sqrt{\abs{g}} \s \lp
\frac{1}{2} \s \pd_t \Phi^T \cdot \pd_t \Phi
- \frac{1}{2} \s \pd_z \Phi^T \cdot \pd_z \Phi
\right. \nonumber \\ & \hspace*{0.5\linewidth} \left.
- \frac{1}{2} \s \Phi^T \cdot K \cdot \Phi
\rp ,
\end{align}
where $K$ is symmetric for the $L^2$ scalar product, independent of
$(t,z)$, and does not involve $(\pd_t, \pd_z)$, then the linear equation
on perturbations reads
\begin{equation}
\pd_t^2 \Phi - \pd_z^2 \Phi = - K \cdot \Phi.
\end{equation}
The same argument as above (with eventually a minus sign) shows that if
$\pd_t^2 - \pd_z^2$ has a strictly negative (respectively strictly
positive) eigenvalue for the ``restricted'' problem, then it also has a
strictly negative (resp. strictly positive) eigenvalue for the full
problem, with a larger or equal absolute value.

\subsection{Application to the problem studied in the main text}

For simplicity, we work with the neutral model $e_2 = 0$. We look for
solutions where $\A_0 = \A_3 = 0$ and assume that the metric reads
\begin{equation}
\dd s^2 = \dd t^2 - \dd r^2 - r^2 \s \dd \theta^2 - \dd z^2.
\end{equation}  
The Lagrangian density then becomes
\begin{widetext}
\begin{equation} 
\mathcal{L} = 
	- \frac{1}{2} \s \F_{0 i} \s \F^{0 i}
	- \frac{1}{2} \s \F_{3 i} \s \F^{3 i}
	+ \frac{1}{2} \s \abs{\pd_t \phi}^2
	- \frac{1}{2} \s \abs{\pd_z \phi}^2
	+ \frac{1}{2} \s \abs{\pd_t \sigma}^2
	- \frac{1}{2} \s \abs{\pd_z \sigma}^2
	- \frac{1}{4} \s \F_{i j} \s \F^{i j}
	+ \frac{1}{2} \s (D_i \phi) \s (D^i \phi)^*
	+ \frac{1}{2} \s (\pd_i \sigma) \s (\pd^i \sigma)^*
	- V(\phi, \sigma),
\end{equation}
where the indices $i$ and $j$ run from $1$ to $2$. Using that $\A_0 =
\A_3 = 0$, this may be rewritten as
\begin{eqnarray}
\mathcal{L} & = & 
	- \frac{1}{2} \s (\pd_t \A_i) \s (\pd_t \A^i)
	+ \frac{1}{2} \s (\pd_z \A_i) \s (\pd_z \A^i)
	+ \frac{1}{2} \s \abs{\pd_t \phi}^2
	- \frac{1}{2} \s \abs{\pd_z \phi}^2
	+ \frac{1}{2} \s \abs{\pd_t \sigma}^2
	- \frac{1}{2} \s \abs{\pd_z \sigma}^2 \nonumber \cr
	& & - \frac{1}{4} \s \F_{i j} \s \F^{i j}
	+ \frac{1}{2} \s (D_i \phi) \s (D^i \phi)^*
	+ \frac{1}{2} \s (\pd_i \sigma) \s (\pd^i \sigma)^*
	- V(\phi, \sigma).
\end{eqnarray}
To go further, we restrict to solutions of the form
\begin{equation}\label{eq:formphisoogma}
\lb 
\begin{aligned}
& \phi: (t, r, \theta, z) \mapsto \e^{\ii \s n \s \theta} \s \varphi(t, r, z), \\
& \sigma: (t, r, \theta, z) \mapsto \e^{\ii \s (\om \s t - k \s z)} \s \xi(t, r, z),
\end{aligned}
\right.
\end{equation}
where $\varphi$ and $\xi$ are real-valued functions. 
The Lagrangian density may be rewritten as
\begin{align*} 
\mathcal{L} = & \; 
	- \frac{1}{2} \s (\pd_t \A_i) \s (\pd_t \A^i)
	+ \frac{1}{2} \s (\pd_z \A_i) \s (\pd_z \A^i)
	+ \frac{1}{2} \s (\pd_t \varphi)^2
	- \frac{1}{2} \s (\pd_z \varphi)^2
	+ \frac{1}{2} \s (\pd_t \xi)^2
	- \frac{1}{2} \s (\pd_z \xi)^2
	\\ &
	+ (\om^2 - k^2) \s \xi^2
	- \frac{1}{4} \s \F_{i j} \s \F^{i j}
	+ \frac{1}{2} \s (\pd_i \varphi) \s (\pd^i \varphi)
	- \frac{e_1^2}{2 \s r^2} \s (\A_\theta + n)^2 \s \varphi^2
	- \frac{e_1^2}{2} \s \A_r^2 \s \varphi^2
	+ \frac{1}{2} \s (\pd_i \xi) \s (\pd^i \xi)
	- V(\varphi, \xi)
.
\end{align*}
Considering perturbations $\delta \A_i$ of $\A_i$, $\delta \varphi$ of
$\varphi$, and $\delta \xi$ of $\xi$ from a stationary solution
independent of $(t,z)$. The second-order Lagrangian density is:
\begin{align*} 
\mathcal{L}^{(2)} = & \; 
	- \frac{1}{2} \s (\pd_t \delta \A_i) \s (\pd_t \delta \A^i)
	+ \frac{1}{2} \s (\pd_z \delta \A_i) \s (\pd_z \delta \A^i)
	+ \frac{1}{2} \s (\pd_t \delta \varphi)^2
	- \frac{1}{2} \s (\pd_z \delta \varphi)^2
	+ \frac{1}{2} \s (\pd_t \delta \xi)^2
	- \frac{1}{2} \s (\pd_z \delta \xi)^2
	\\ &
	+ (\om^2 - k^2) \s \delta \xi^2
	- \frac{1}{4} \s \delta \F_{i j} \s \delta \F^{i j}
	+ \frac{1}{2} \s (\pd_i \delta \varphi) \s (\pd^i \delta \varphi)
	- \frac{e_1^2}{2 \s r^2} \s (\A_\theta + n)^2 \s (\delta \varphi)^2
	- \frac{e_1^2}{2 \s r^2} \s \delta \A_\theta^2 \s \varphi^2
	- \frac{e_1^2}{r^2} \s (\A_\theta + n) \s \varphi \s \delta \A_\theta \s \delta \varphi
	\\ & 
	- \frac{e_1^2}{2} \s (\delta \A_r)^2 \s \varphi^2
	- \frac{e_1^2}{2} \s \A_r^2 \s (\delta \varphi)^2
	- e_1^2 \s \A_r \s \varphi \s \delta \A_r \s \delta \varphi
	+ \frac{1}{2} \s (\pd_i \delta \xi) \s (\pd^i \delta \xi)
	\\ &
	- \frac{1}{2} \s \pd_\varphi^2 V(\varphi, \xi) \s (\delta \varphi)^2
	- \frac{1}{2} \s \pd_\xi^2 V(\varphi, \xi) \s (\delta \xi)^2
	- \pd_\varphi \pd_\xi V(\varphi, \xi) \s \delta \varphi \s \delta \xi
,
\end{align*} 
where $\delta \F_{i,j} \equiv \pd_i \delta \A_j - \pd_j \delta \A_i$.
Let us define
\begin{equation}
\Phi \equiv 
\begin{pmatrix}
\delta \xi \\
\delta \varphi \\
\delta \A_r \\
\delta \A_\theta / r
\end{pmatrix} .
\end{equation}
We obtain:
\begin{equation}
\mathcal{L}^{(2)} = 
	\frac{1}{2} (\pd_t \Phi)^T \cdot (\pd_t \Phi)
	- \frac{1}{2} (\pd_z \Phi)^T \cdot (\pd_z \Phi)
	- \frac{1}{2} \s \Phi^T \cdot K \cdot \Phi
	+ \cdots,
\end{equation}
\end{widetext}
where ``$\cdots$'' denotes total derivatives obtained by integration by
parts to make all derivatives in $r$ and $\theta$ act on $\Phi$, and $K$
is a differential operator involving only $\pd_i$ and depending only on
$r$ and $\theta$. The action may thus be written in the
form~\eqref{eq:formQ}.

\bibliographystyle{apsrev4-1}
\bibliography{references}

\end{document}